\documentstyle[12pt,epsf,cite]{article}
\begin{document}
\def\be{\begin{equation}}
\def\ee{\end{equation}}
\def\barr{\begin{array}}
\def\earr{\end{array}}
\def\bea {\begin{eqnarray}}
\def\eea {\end{eqnarray}}
\def\benum{\begin{enumerate}}
\def\eenum{\end{enumerate}}
\def\bitem{\begin{itemize}}
\def\eitem{\end{itemize}}
\def\dis{\displaystyle}

\def\ra {\rightarrow}
\def\lra {\longrightarrow}
\def\bra{\langle}
\def\ket{\rangle}
\def\lsim{\:\raisebox{-0.5ex}{$\stackrel{\textstyle<}{\sim}$}\:}
\def\gsim{\:\raisebox{-0.5ex}{$\stackrel{\textstyle>}{\sim}$}\:}
\def\lappeq{\mathrel{\rlap{\raise.5ex\hbox{$<$}}
                    {\lower.5ex\hbox{$\sim$}}}}
\def\hrar{\hookrightarrow}
\def\bul{\bullet}
                              \def\ptsl{p_T \hspace{-1.1em}/\;}
                              \def\gev{\: \rm GeV} 
                              \def\tev{\: \rm TeV} 
                              \def\pb {\: \rm pb}
                              \def\fb {\: \rm fb}
\def\RedOrange{} 
\def\RedViolet{} 
\def\Black{}
\def\Blue{}
\def\JungleGreen{}
\def\RawSienna{}
\newcommand{\no}{\!\!\!\!/}
\def\G0{\widetilde G}
\def\N0{\widetilde \chi^0}
\def\Cp{\widetilde \chi^+}
\def\Cm{\widetilde \chi^-}
\def\Cpm{\widetilde \chi^\pm}
\def\Cmp{\widetilde \chi^\mp}
\def\sq{\widetilde q}
\def\su{\widetilde u}
\def\sd{\widetilde d}
\def\sc{\widetilde c}
\def\ss{\widetilde s}
\def\st{\widetilde t}
\def\sb{\widetilde b}
\def\sl{\widetilde \ell}
\def\se{\widetilde e}
\def\er{\widetilde e_R}
\def\snu{\widetilde \nu}
\def\smu{\widetilde \mu}
\def\stau{\widetilde \tau}
\def\tm{\widetilde m}
\def\cone {\chi_1^0}
\def\ctwo {\chi_2^0}
\def\mc {m_{\chi_1^0}}
\def\L {\Lambda}
\def\s {\sigma}
\def\grav {\tilde G}
\def\tb{\tan \beta}
\def\rbf#1{{\RedOrange \bf #1}\Black \rm}
\def\bluebf#1{{\Blue \bf #1}\Black \rm}
\def\pinegbf#1{{ \bf #1}\Black \rm}
\def\mahogbf#1{{ \bf #1}\Black \rm}
\def\mult{\multicolumn{1}{|c|}}
\def\multgev{\mult{(GeV)}}
\def\multtev{\mult{(TeV)}}
%
\def\rpv{$R_p \hspace{-1em}/\;\:$}
\def\rp{$R$-parity}
\def\l{\lambda}
\def\lp{\lambda'}
\def\lpp{\lambda''}

                         \def\eg{ {\em e.g.}~}
                         \def\etal{ {\em et al.}}
                         \def\ie{ {\em i.e.}~}
                         \def\viz{ {\em viz.}~}

\def\mET{E_T \hspace{-1.1em}/\;\:}
\def\mpT{p_T \hspace{-1em}/\;\:}

\def\epem{e^+ e^-}
\def\emem{e^- e^-}
\def\bib {\bibitem}

\def\a{\alpha}
\def\as {\alpha_s}
\def\b{\beta}
\def\g{\gamma}
\def\d{\delta}
\def\e{\epsilon}
\def\ve{\varepsilon}
\def\l{\lambda}
\def\m{\mu}
\def\n{\nu}
\def\G{\Gamma}
\def\D{\Delta}
\def\L{\Lambda}
\def\s{\sigma}
\def\p{\pi}

\def\t {\times }
\def\slash {\!\!\!\!\!\!/}
\def\photino {\tilde\gamma}
\def\neu {\chi_1^0}
\def\rslep {\tilde e_R}
\def\lslep {\tilde e_L}
\def\mrslep {m_{\rslep}}
\def\mlslep {m_{\lslep}}
\def\mneu {m_{\neu}}

\thispagestyle{empty}
\begin{flushright}
						    MRI-P-000429\\
                                                   TIFR-TH/00-35  \\
                                                   {\large \tt hep-ph/0007139}
\end{flushright}

\vspace*{1in}

\begin{center}

{\LARGE\bf Signals for Gauge Mediated Supersymmetry \\ \bigskip 
        Breaking Model at an {${\bf e}^{\bf -}{\bf e}^{\bf -}$} Collider}\\

\vskip 25pt

{\bf                 Debajyoti Choudhury } \\ 

{\footnotesize\rm 
                Mehta Research Institute,\\
                Chhatnag Road, Jhusi, \\
                Allahabad - 211 019, India\\
                E-mail: debchou@mri.ernet.in } \\

\bigskip

{\bf                        Dilip Kumar Ghosh} \\ 

{\footnotesize\rm 
                       Department of Theoretical Physics,\\
                       Tata Institute of Fundamental Research, \\
                       Homi Bhabha Road,\\
                        Mumbai 400 005, India. \\ 
                     E-mail: dghosh@theory.tifr.res.in  } \\ 

\vskip 30pt

{\bf                             Abstract 
}

\end{center}

\begin{quotation}
\noindent
We study the pair-production and decay of right handed selectrons 
within a  Gauge Mediated Supersymmetry Breaking Model in polarised 
$e^-e^-$ interaction. Detailed analyses of the possible signals and 
backgrounds are performed for a few selected points in the parameter 
space. A judicious choice for polarisation of the initial electron 
beam helps eliminate almost the entire Standard Model background. 
We also show that phase space distributions can be used to distinguish
such a supersummetry breaking scheme from the supergravity inspired 
models.

\end{quotation}

\vskip 60pt


\newpage

\section{Introduction}

The Standard Model (SM) of high energy physics, despite its eminent 
success, suffers from certain drawbacks, the hierarchy problem 
being a major one. Supersymmetry (SUSY) provides an elegant solution
and, consequently, has been a cornerstone in attempts to build models 
going beyond the SM. It is manifestly clear, though, that SUSY must 
be broken, at least at low energies. This forces us onto a 
new problem. Since SUSY cannot be broken in the observable sector
in a phenomenologically consistent way~\cite{supertrace}, 
one is forced to introduce a hidden sector wherein the breaking 
takes place. The question as to how this breaking is conveyed to the 
observable sector is yet to be settled. The idea that gravity plays 
the primary mediating role~\cite{sugra} has, 
historically, been the most popular one. In such 
supergravity (SUGRA)-inspired models, SUSY breaking 
occurs at a very high scale (typically well above
the grand unification scale) and is communicated to the visible sector 
through gravitational interactions, the only one common to both the 
sectors. The gravitino turns out be heavy (at or above the electroweak 
scale) and, generically, 
the lightest of the neutralinos is the lightest supersymmetric particle 
(LSP). Such scenarios suffer from a potential drawback 
though: interactions of heavy fields above the Grand Unification scale 
can induce large flavour-changing neutral currents at low 
energies~\cite{hall}. Questions like these as well as the fact 
that we still do not have a complete theory of SUSY breaking through 
gravitation have, in recent years, prompted research into 
alternative mechanisms
~\cite{gmsb_old,gmsb_gut,gmsb2,gmsb3,gmsb4,gmsb_review,AMSB_model,AMSB_pheno} 
for SUSY breaking.

One such mechanism postulates a set of particles (the ``messenger sector''
MS) that both transform non-trivially under 
the SM gauge group, as well as interact with the hidden sector. 
The latter interaction, which communicates SUSY breaking from the 
hidden sector to the MS superfield(s), could have a 
characteristic scale as low as ${\cal O}(10^{2-3}) \tev$~\cite{gmsb_review}. 
The SM 
gauge interactions can then serve to communicate the breaking to the 
observable sector. This assures, for example, that the MSSM 
sfermions with the same quantum numbers are degenerate at the 
scale. Furthermore, given the limited range of renormalization group 
running, they continue to be approximately degenerate at the electroweak 
scale thereby avoiding the flavour problem. Even more interestingly, 
the gravitino in these gauge mediated SUSY breaking (GMSB) models 
turns out to be superlight, in contrast to the case of the 
supergravity models. 
Consequently, the lightest of the usual superpartners (now 
the next to lightest supersymmetric particle or NLSP) can now decay 
into its SM counterpart and the gravitino. 

We see thus, that, apart from its purely theoretical aspects,
the dynamics of SUSY breaking is likely to leave its imprint on 
low energy phenomenology as 
well~\cite{gmsb_teva_lep,gmsb_teva,gmsb_lep,gmsb_nlc,hontka}. With 
the spectrum changing significantly, search strategies need to be 
modified. Furthermore, there could be cases where, even after SUSY 
signals have been established, an understanding of the mode of SUSY 
breaking remains elusive~\cite{hontka}. Such ``inadequacies'' of the 
simplest strategies thus call for new ones to be developed. We shall 
attempt to do this in the context of $\emem$ coliders. 

We structure the rest of this article as follows. In Section~\ref{sec:GMSB},
we present a very brief review of the GMSB models. The following section 
deals with selectron pair production at $\emem$ colliders. 
In sections~\ref{sec:selec_NLSP} and \ref{sec:neut_NLSP}, we 
examine the signal and background for cases with selectron NLSP 
and neutralino NLSP respectively. Section~\ref{sec:identify} 
examines the possibility of identifying between GMSB and SUGRA-inspired 
models. Finally, we conclude in Section~\ref{sec:concl}.

\section{The spectrum in GMSB models}
	\label{sec:GMSB}
Renormalizability of the theory, coupled with economy of field content,
dictates that the messenger sector 
be comprised of chiral superfields such that their SM gauge 
couplings are vectorial in nature. Most GMSB models actually consider 
these fields to be in 
($5 + \bar 5$) or ($10 + \overline{10}$) representations of $SU(5)$. This 
construct, while not mandatory, helps preserve the successful 
SUSY-GUT prediction of the weak mixing angle. 
The maximum number of messenger families is constrained by the twin 
requirements of low energy supersymmetry breaking 
and perturbativity upto the grand unification scale
to five of ($5 + \bar 5$)s or to one ($10 + \overline{10}$) in addition 
to two pairs of ($5 + \bar 5$).

Restricting ourselves, for the time being, to a single pair of 
MS supermultiplets ($\Psi + \bar \Psi$), consider a term in the 
superpotential of the form $\lambda {\cal S} \bar \Psi \Psi$, where 
${\cal S}$ is an SM singlet. The scalar ($S$) and auxilliary ($F_S$)
components of ${\cal S}$ may acquire vacuum expectation values (vevs)
through their interactions with the hidden sector fields. SUSY breaking 
is thus communicated to the MS, with the fermions and sfermions 
acquiring different masses. This, in turn, is communicated to the 
SM fields resulting in the gauginos and sfermions acquiring masses 
at the one-loop and two-loop levels respectively. The expressions,  
in the general case of multiple messenger pairs and/or gauge singlets
${\cal S}_i$, is a somewhat 
complicated function~\cite{gmsb_review} of $M \equiv \bra S \ket$ and 
$\L \equiv \bra F_S \ket / \bra S \ket$. However if there be just one such 
singlet, the expressions for masses at the messenger scale $M$ 
simplify to
\be
\barr{rcl}
{\tilde M}_i(M) &=& \dis 
      N_m \: \frac{\alpha_i(M)}{4\pi} \: \L \: 
                   f_1 \left(\frac{\L}{M}\right)  \\[2ex]
{\tilde m}^2_{\tilde f}(M) &=& \dis 
	2 N_m \: \L^2  f_2 \left(\frac{\L}{M} \right) \: 
              \sum^3_{i=1}\kappa_i C^{\tilde f}_i 
			\left(\frac{\alpha_i(M)}{4\pi}\right)^2 \ .
	\label{threshold}
\earr
\ee
where $N_m$ is the number of messenger generations.
In eq.(\ref{threshold}), $ C^{\tilde f}_i$ are the quadratic Casimirs 
for the sfermion in question. The factors $\kappa_i$ equal 1, 1 and 
$5/3$ for $SU(3)$, $SU(2)$ and $U(1)$ respectively with the 
gauge couplings so normalized that $\kappa_i \alpha_i$ are equal 
at the messenger scale. The threshold functions are given by
\bea
f_1(x) &=& \frac{1+x}{x^2} \log(1+x) + \big (x \ra -x \big)\\
f_2(x) &=& \frac{(1+x)}{x^2}
	\left[\log(1+x) + 2 Li_2\left(\frac{x}{1+x}\right) 
	                 - \frac{1}{2}Li_2\left(\frac{2x}{1+x}\right)
	\right] + \big (x \ra -x \big) \ .
\eea
The superparticle masses at the electroweak scale are obtained 
from those in eq.(\ref{threshold}) by evolving the appropriate
renormalization group equations. For the scalar masses, the $D$-terms 
need to be added too.

\section{Selectron production in polarized $\emem$ colliders} 
	\label{sec:prodn}

\begin{figure}[htb]
\centerline{
\epsfxsize=8cm\epsfysize=7cm
                    \epsfbox{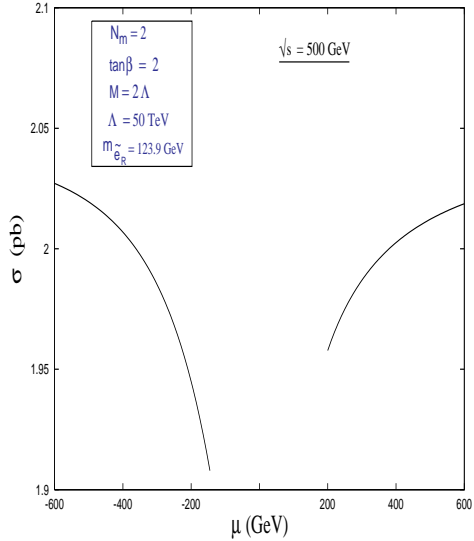}
	   }
\caption{\em The dependence of $\rslep$ pair production 
	     cross section on the higgsino mass parameter $\mu$. 
	     For the assumed set of parameters, the region
	     $|\mu| \lappeq 200$ GeV has already been ruled 
	     out by the existing lower limits on  $ \N0_1$ and $\Cp_1$ 
             masses~\protect\cite{ALEPH}.
        } 
       \label{fig:prod_mu}
\end{figure}

At an $\emem$ collider, the dominant production mode for supersymmetric 
particles is that of a pair of selectrons~\cite{keung,frank}. The relevant
term in the Lagrangian reads
\[
  {\cal L} = e_{\rm em} \bar \chi^0_a (l_{i a} P_L + r_{i a} P_R) e \; 
		\tilde \e_i \ ,
\]
where $\chi^0_a$ ($a = 1 \dots 4$) represent the neutralino fields and
$e_i$ refer to $\tilde e_{L, R}$ as the case may be. Consider 
the polarized electron scattering 
\be
    e^-(\lambda_1) + e^-(\lambda_2)
       \longrightarrow  \se_1^-(m_1) + \se_2^-(m_2)
\ee
which proceeds through the 
$t$- and $u$-channel exchanges of each of the four neutralinos.
The corresponding 
differential cross sections are given by
\be
\barr{rcl}
\dis \frac{ {\rm d} \sigma}{ {\rm d} t } 
     & = & \dis
       \frac{\pi \alpha^2}{4 s} \sum_{a,b} 
              \Bigg[  s M_a M_b D_{a b} 
                     \left( \frac{1}{t - M_a^2} + \frac{1}{u - M_b^2}
                     \right)^2
           - 2 I_{a b} \;
                   \frac{(u t - m_1^2 m_2^2)}{(t - M_a^2) (u - M_b^2)} 
         \\[3ex]
      & & \dis \hspace*{3em}
           + (u t - m_1^2 m_2^2) C_{a b} 
                   \left( \frac{1}{(t - M_a^2) (t - M_b^2)} + 
                          \frac{1}{(u - M_a^2) (u - M_b^2)}
                     \right)
       \Bigg]
     \\[4ex]
D_{ab} & \equiv & \dis
             l_{1 a} l_{1 b} l_{2 a} l_{2 b} 
                             (1 - \lambda_1) (1 - \lambda_2)
           + r_{1 a} r_{1 b} r_{2 a} r_{2 b} 
                             (1 + \lambda_1) (1 + \lambda_2)
     \\[4ex]
C_{ab} & \equiv & \dis
             l_{1 a} l_{1 b} r_{2 a} r_{2 b} 
                             (1 + \lambda_1) (1 - \lambda_2)
           + r_{1 a} r_{1 b} l_{2 a} l_{2 b} 
                             (1 - \lambda_1) (1 + \lambda_2)
     \\[4ex]
I_{ab} & \equiv & \dis
             l_{1 a} r_{1 b} r_{2 a} l_{2 b} 
                             (1 + \lambda_1) (1 - \lambda_2)
           + r_{1 a} l_{1 b} l_{2 a} r_{2 b} 
                             (1 - \lambda_1) (1 + \lambda_2) \ ,
\earr
	\label{prod_cs}
\ee
where $M_a$  are the masses of the neutralinos.
The masses, as well as the couplings $l_{ia}$ and $r_{ia}$, 
are, of course, determined by $N_m$, $\L$, $M$ as well as 
$\tb$ and $\mu$. 
Since the coupling of a fermion-sfermion pair to a higgsino 
is proportional to the fermion mass, clearly the higgsino components 
of the neutralinos play only a small part in selectron production. 
In other words, the dependence of the cross section
on $\mu$ (and $\tb$) is of a minor 
nature (see Fig.~\ref{fig:prod_mu}). As far as decays of
the selectron are concerned, these parameters
do play a more significant role though.
A small $\mu$, for example, results in the some of the 
neutralinos being light, thus 
affording a decay channel which might not be otherwise available to 
the selectron. Inspite of the small coupling of the electron-selectron 
pair to the higgsinos, such decays might be competitive with the gravitino 
mode. However, since we would be explicitly considering 
such cascading decays of the selectron, we are justified in neglecting
the dependence on $\mu$ and $\tb$.
%

\begin{figure}[ht]
\centerline{
\epsfxsize=7cm\epsfysize=7cm
                     \epsfbox{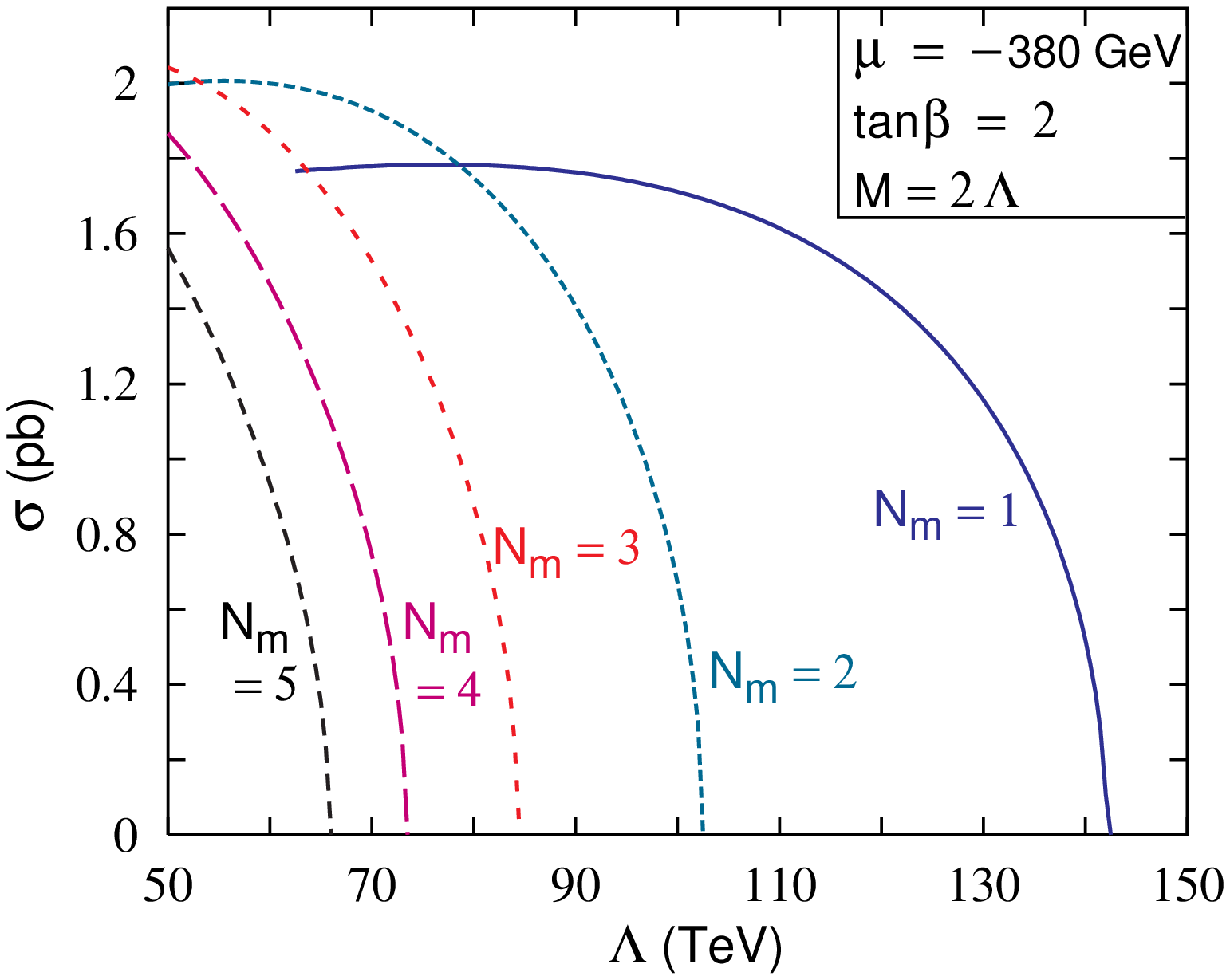}
\epsfxsize=7cm\epsfysize=7cm
                     \epsfbox{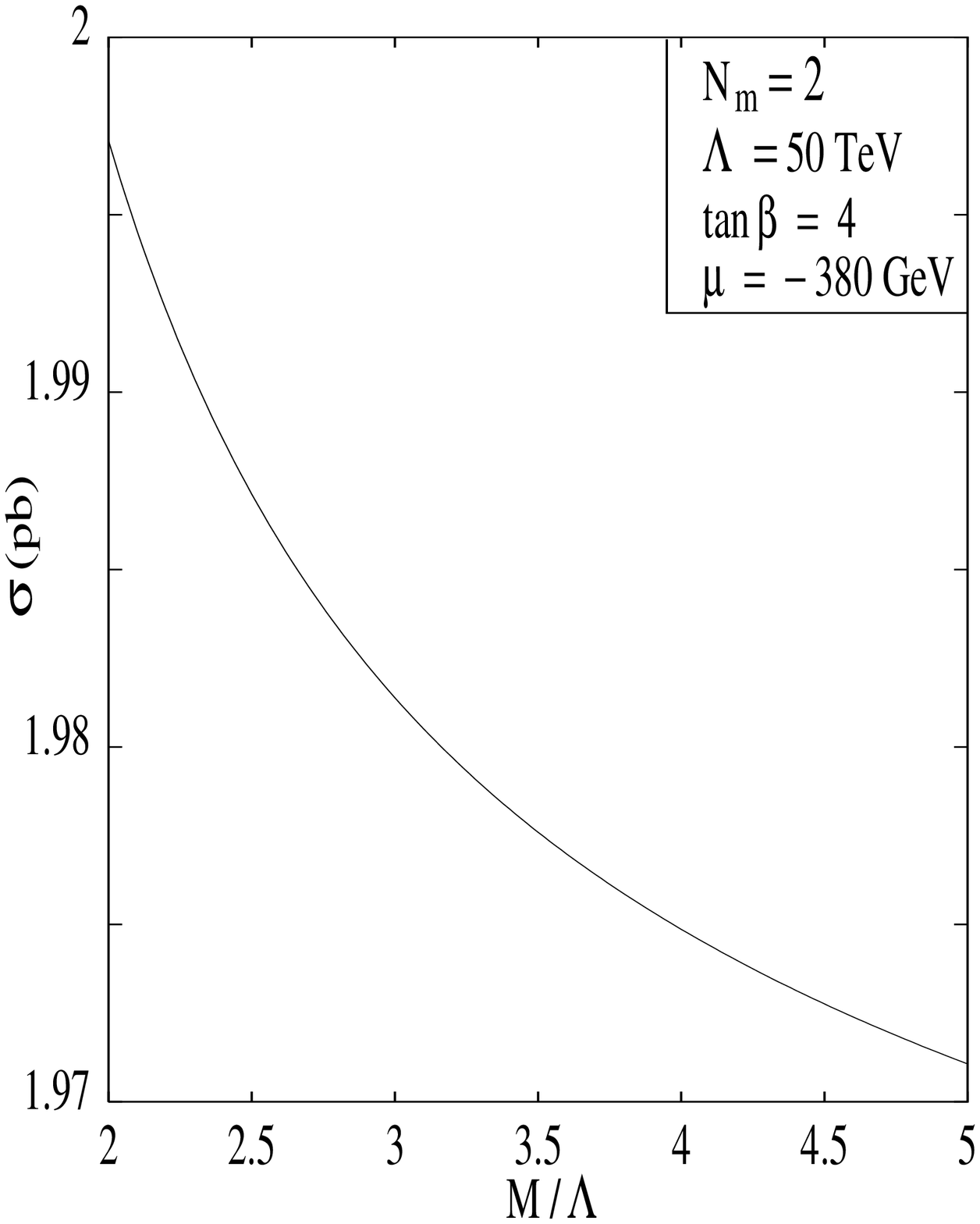}
}
\caption{\em {\em (a)} The dependence of $\rslep$ pair production 
	     cross section (for $\sqrt s = 500 \gev$) 
	     on the scale $\L$ for a fixed ratio 
	     of $M/ \L$ and given parameters. The intercepts on the 
	     $\L$-axis essentially denote the kinematic threshold.
	     For $N_m = 1$, 
	     $\L < 60 \gev$ leads to a spectrum
	     inconsistent with existing lower limits on 
             $ \N0_1$ and $\Cp_1$ masses~\protect\cite{ALEPH}.
	     {\em (b)} The dependence of the cross section 
	     on the ratio $M / \L$ for given $\L$.
        }
    \label{fig:prod_lambda}
\end{figure}

For a given ratio $\L / M$, both the scalar and the gaugino masses grow 
with $\L$ (eq.~\ref{threshold}). Consequently, the production cross section
falls steeply with $\L$ (see Fig.~\ref{fig:prod_lambda}$a$). 
The fall is understandably steeper 
for larger $N_m$ as the selectron mass grows as $\L \sqrt{N_m}$. 
The behaviour for small $\L$ is more subtle. The total cross section
is a complicated function of $M_a / \sqrt{s}$ and $m_{\se} / \sqrt{s}$. 
Combined with the fact that the couplings with the $\widetilde B$ 
and $\widetilde W_3$ are different, this can lead to a situation where 
the cross sections do not actually fall with $N_m$ 
(Fig.~\ref{fig:prod_lambda}$a$). On the other hand,
as the mass parameter $M$ enters eq.(\ref{threshold}) only logarithmically, 
one may expect that the cross-sections would change only fractionally 
as this parameter is varied (for a fixed $\L$). This is borne out by 
Fig.~\ref{fig:prod_lambda}$b$. Since such deviations are almost of the 
order of statistical fluctuations in the signal itself, for
the rest of our study, we will not consider any explicit 
dependence on the mass ratio $M / \L$.

As we have already pointed out, in GMSB models, 
the $\lslep$ is distinctly heavier than the $\rslep$. Hence, 
we shall concentrate on the pair-production process 
$\emem \ra \rslep \rslep$. Once produced, the selectron may decay 
to either $e^- + \grav$ or $e^- + \neu$ (if kinematically allowed). 
In the first case the final state comprises of two electrons and 
missing momentum, while the second case has two photons in addition. 
As the backgrounds are quite different, we shall now examine 
each case individually.

\section{Selectron as the NLSP}
	\label{sec:selec_NLSP}
With the selectrons decaying into an electron and a gravitino each, 
the SM background comprises of $\emem \nu_i \bar \nu_i$. The main 
contributions to the latter clearly arise from the ``resonant'' 
processes $\emem \lra e^- \nu_e W^-$ and $\emem \lra \emem Z$ and have 
been discussed at some length in refs.~\cite{frank,frank_dc}. The 
$W$ contribution is dominant but can be suppressed by right-polarizing 
the electron beams. This also serves to enhance the selectron production 
rate. Of course, the ideal state of a fully polarized beam 
is virtually unattainable and hereon 
we shall assume the electron beams to be 90\% right-polarized. 

It is obvious that the phasespace distribution of the signal events 
would depend crucially on the mass of the selectron and, to a lesser 
extent, on the neutralino masses. For purposes of comparison between 
the signal and the background, we will concentrate on two 
specific choices in the parameter space marked in Table~\ref{table:sel_cs}.
\begin{table}[htb]
\begin{center}
\begin{tabular}{|r|r|r|r|r|r|r|r|r|}
\cline{2-9}
\multicolumn{1}{c|}{}
& \mult{\bf $N_m$} & \mult{\bf $M $} & \mult{\bf $ \Lambda $  } & 
	\mult{\bf $\mu$} & \mult{\bf $\tan\beta $} & 
        \mult{\bf $ m_{\tilde e_{R} } $} &  \mult{{\bf $\sigma $} (fb)} & 
        \mult{{\bf $\sigma $} (fb)}
	\\ 
\multicolumn{1}{c|}{}
&   &  \multtev  & \multtev & \multgev
&    &  \multgev & \mult{$\sqrt{s}=0.5$~TeV} & \mult{$\sqrt{s}=1$~TeV}\\
\hline
\rbf{(A)}& \rbf{2} &  \rbf{100}  & \rbf{50.0}  
		& \rbf{--500}  & \rbf{2} &  \rbf{123.9} 
		& \rbf{830.3} & \rbf{37.91}\\
\hline
& 2 &  120  & 60.0  & 400  & 2 &  147.3 &  833.2 & 43.18\\
\hline
& 2 &  200  & 100.0 & 400  & 3 &  243.8 &  202.7 & 37.99\\
\hline
& 3 &  100  &50.0 &$-400$  & 2 &  149.1  &  851.5 &44.96 \\
\hline
&  3 &  150  &75.0 & 350  & 4 &  222.8 &  483.2 & 42.42\\
\hline
& 3 &  250  &62.5 & $-350$  & 4 &  188.9 & 705.3 &42.41 \\
\hline
& 4 &  100  & 50.0 & $-250$  & 2 &  170.5 &  766.5 &46.63 \\
\hline
& 4 &  250  & 62.5 & 450  & 4 &  216.2 &  519.1 &43.72 \\
\hline
\bluebf{(B)} & \bluebf{4} &  \bluebf{300}  & \bluebf{60.0}  & \bluebf{--450} 
	& \bluebf{4} &  \bluebf{208.6} &  \bluebf{581.7} & \bluebf{44.27} \\
\hline
& 5 &  100  & 50.0  & $-500$ & 2 &  189.6 &  643.4 & 46.93\\
\hline
& 5 &  250  & 62.5 & 300 & 3 &  240.0 &  214.3& 40.52\\
\hline
& 5 &  300  & 60.0 & $-250$ & 4 &  231.9&  320.2& 41.98\\
\hline
\end{tabular}
\end{center}
\caption {\em Signal ($ e^- e^- + \mpT $ ) cross-section
          for some representative values of GMSB input parameters.
	  The cuts of eqs.(\protect\ref{eta_e}--\protect\ref{miss_mom}) 
	  have been imposed. 
         }
	\label{table:sel_cs}
\end{table}

Whereas the electrons from an $\rslep$ would be produced isotropically, 
the background events would prefer to have at least one of them to be 
close to the beam pipe~\cite{frank_dc}. We thus demand that both 
the electrons must satisfy
\be
	| \eta_e | < 3 \ ,
	\label{eta_e}
\ee	
a requirement in consistency with the angular coverage of proposed 
detectors. In addition, the leptons must have sufficient momentum 
to be detectable, {\em viz}
\be
	p_T(e) > 5 \gev \ ,
	\label{pt_e}
\ee	
and be separated enough to be individually resolved:
\be
	\D R \equiv \sqrt{ (\D \eta)^2 + (\D \phi)^2} > 0.2 \ ,
	\label{delta_R_e}
\ee	
where $\D \phi$ refers to their azimuthal separation. In addition,
we demand that the missing momentum be large enough:
\be
	\ptsl > 20 \gev  \ .
	\label{miss_mom}
\ee

\begin{figure}[htb]
\vspace*{-0.0cm}
\hspace*{-0.5cm}
\centerline{
\epsfxsize=6.cm\epsfysize=6.5cm
\epsfbox{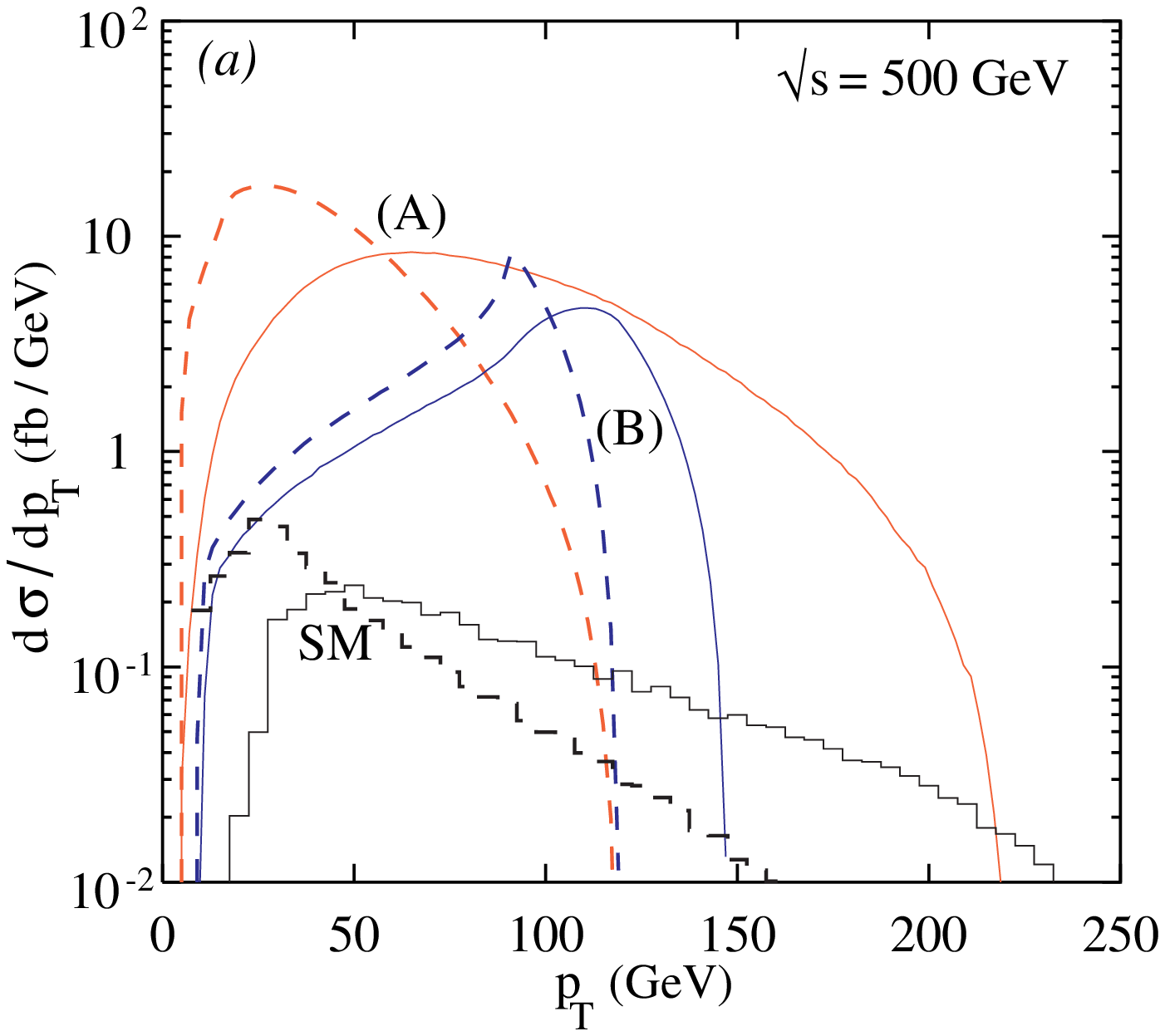} 
\vspace*{-0.0cm}
\hspace*{-0.7cm}
\epsfxsize=6.cm\epsfysize=6.5cm
\epsfbox{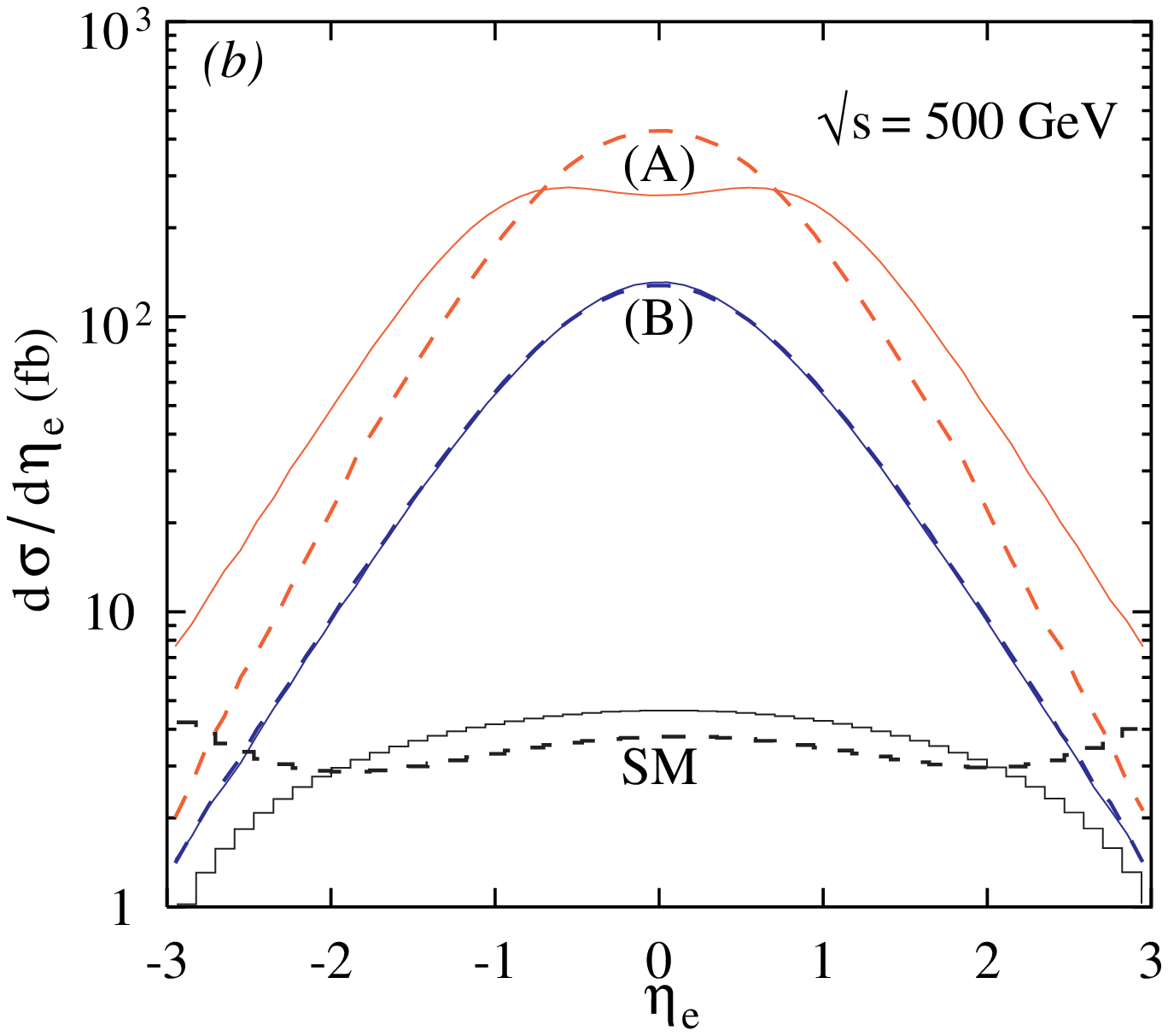}
\vspace*{-0.0cm}
\hspace*{-0.7cm}
\epsfxsize=6.cm\epsfysize=6.5cm
\epsfbox{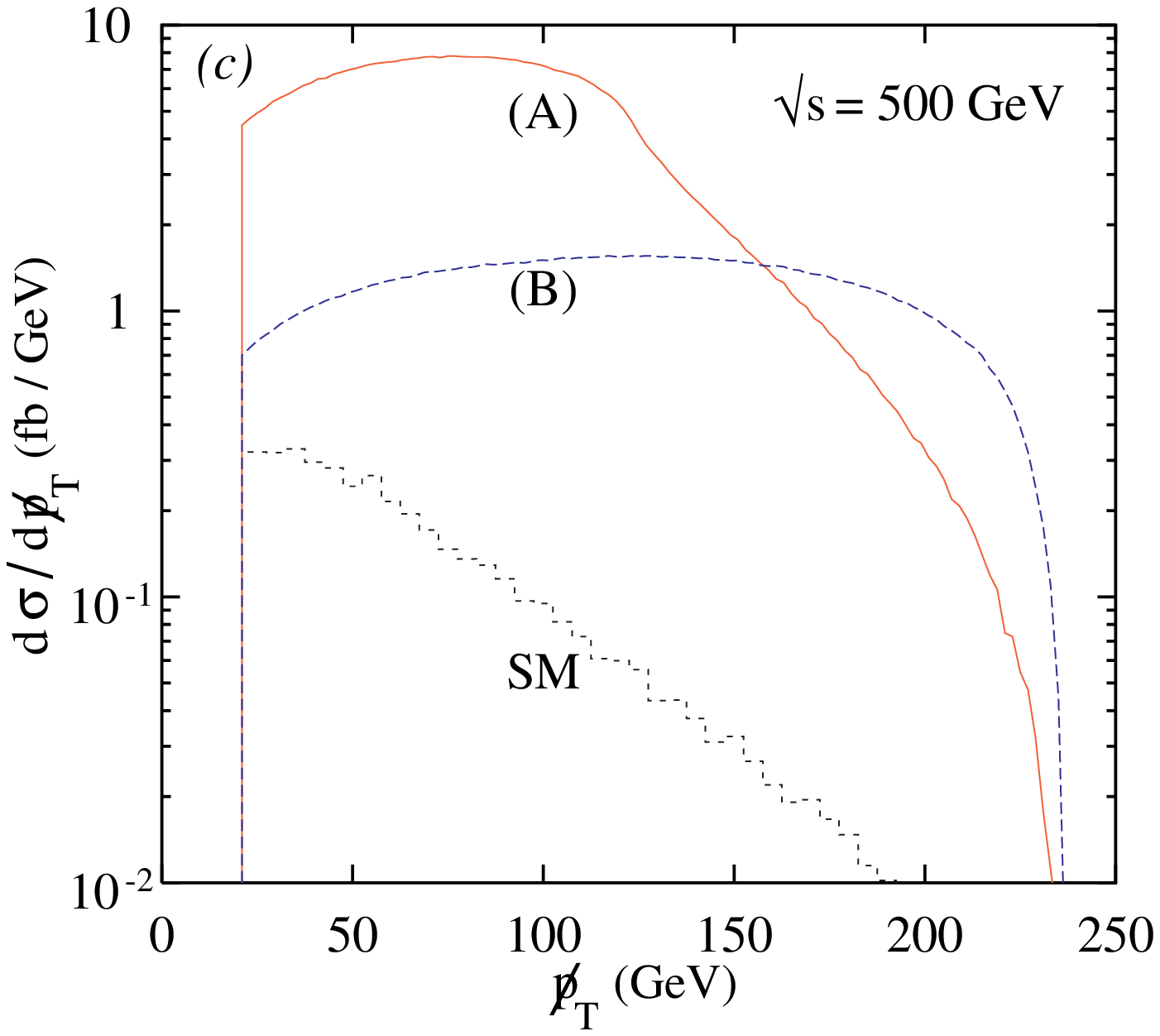}
\vspace*{-0.3cm}
}
\caption{\em Signal and background distributions for the selectron NLSP case. 
	     The histograms correspond to the SM background while the two 
	     sets of smooth curves correspond to the signal 
	     for parameter sets  marked in 
	     Table.~\protect\ref{table:sel_cs}. The cuts of 
	     eqs.(\protect\ref{eta_e}--\protect\ref{miss_mom}) 
	     have been imposed.
	     Graph {\em (a)} pertains to
	     electron transverse momenta with the solid curves 
	     corresponding to the harder electron and the dashed to the 
	     softer one. Similarly, graph {\em (b)} pertains to the 
	     electron rapidities. {\em (c)} 
	     gives the distribution in missinng transverse momentum.
	}
    \label{fig:selec}
\end{figure}

With these cuts in place, the total SM background
is very flat over the region $\sqrt{s} = 500 \gev$--$1 \tev$ and 
amounts to  approximately $19.5 \fb$ over the entire range. 
In Table~\ref{table:sel_cs}, we display the 
signal cross-section for some representative values of GMSB parameters. 
One can get $\sim 10^4$ events per year assuming integrated luminosity
of $50$~${\rm fb}^{-1}$ for $\sqrt{s}= 500 $~GeV and $\sim 10^3$ events for 
$\sqrt{s} = 1$~TeV machine (for the same luminosity). 
In Fig.~\ref{fig:selec}, we present the 
phasespace distribution for the background events. Also shown in the 
figure are the corresponding signal profiles for the two particular points 
in the parameter space. Let us begin by discussing these distributions.

At first sight, it might seem surprising that the transverse momentum 
distribution (see Fig.~\ref{fig:selec}$a$) 
for the signal events do not show the characteristic 
Jacobian peaks. This is not surprising though, as,  for such behaviour
to be exhibited, an electron should only appear as a decay 
product of a {\em particular} particle. However, in the case at hand, the 
two electrons cannot be distinguished from each other and hence have to 
be ordered in some fashion, whether energy or magnitude of transverse 
momentum or rapidity. We choose the first option, namely energy 
ordering. Any such ordering will tend to destroy features indicative
of individual decay product kinematics, and this is particularly true 
of set ($A$). For set ($B$), on the other hand, the selectron mass is 
much closer to $\sqrt{s}/2$ and hence they are produced with very little 
momenta. Consequently, the effect of ordering is relatively smaller 
and the remnant of the Jacobian peak more pronounced.

The rapidity distribution (Fig.~\ref{fig:selec}$b$) 
for the SM background shows clearly that the 
softer electron prefers to lie closer to the beam pipe while the harder 
one is much more central. This is reflective of the singularities 
in the photon-mediated contributions. This also makes itself felt 
in the $\ptsl\ $ 
distributions (see Fig.~\ref{fig:selec}$c$) where the 
SM cross sections fall off much more steeply than those for the 
signal. 

Although the SM background is not too big, it might be desirable  
to reduce it further without sacrificing a large fraction of the 
signal. This becomes particularly important when the signal size 
reduces either on account of $m_{\rslep}$ being close to the kinematic limit 
or other (nonstandard) decay modes becoming available to the 
selectron. A look at Figs.~\ref{fig:selec} tells clearly that this 
goal is unlikely to be achieved by imposing harder cuts on 
either the individual electron $p_T$s or on the missing momentum. 
Removing events with $\eta_e > 2$ is an option, but even then, the 
improvement is merely quantitative and not a qualitative one. 

\begin{figure}[hbt]
\centerline{
\epsfxsize=7cm\epsfysize=7cm
                     \epsfbox{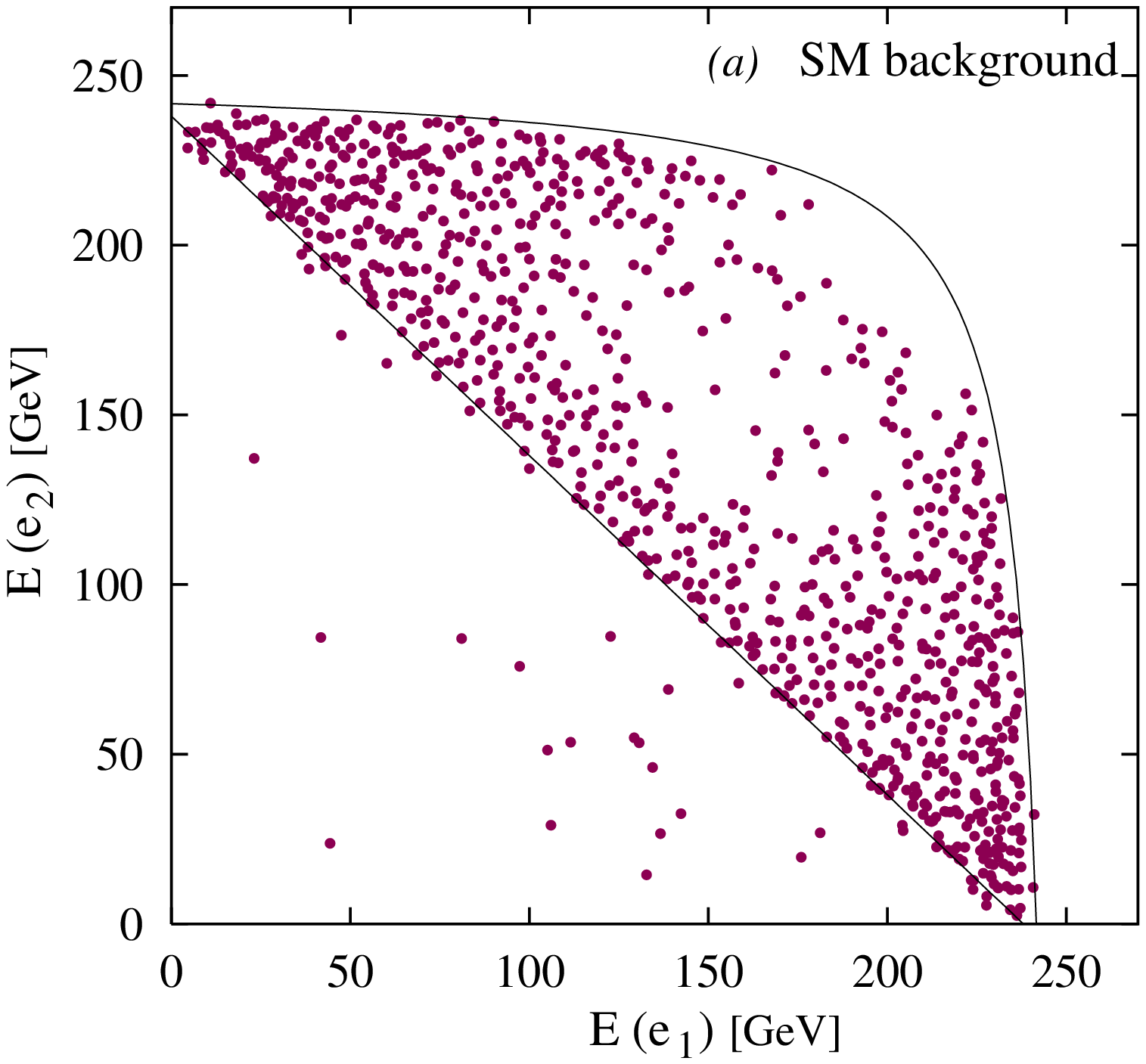}
\epsfxsize=7cm\epsfysize=7cm
                     \epsfbox{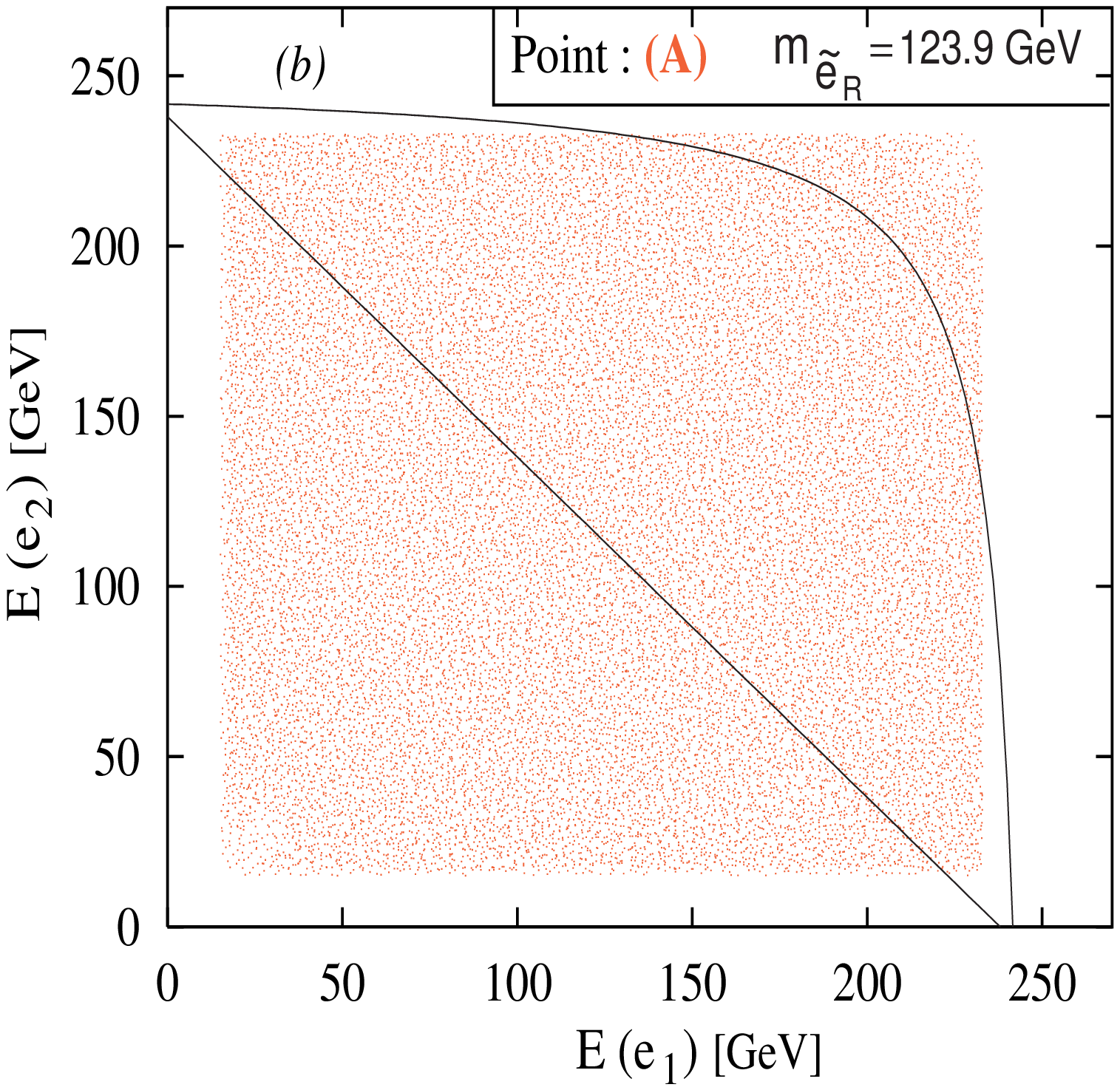}
}
\caption{\em The distribution of the electron energies 
	     for the process $\emem \ra \emem \nu_i \bar \nu_i$
	    for $\sqrt{s} = 500 \gev$ and an integrated luminosity 
	    of $50 \fb^{-1}$. The cuts of 
	    eqs.(\protect\ref{eta_e}--\protect\ref{miss_mom})
	    have been imposed. 
	    {\em (a)} The SM background. 
   	    Each point corresponds to one event. 
	    The events lying in the area enclosed by the straight line 
	    and the curve arise from on-shell $Z$ production 
	    (see eq.\protect\ref{Z_bkgd_limits}). 
	{\em (b)}The signal events 
	    for a particular point in the parameter space. 
	    Each point corresponds to {\bf four} events. The 
	    edges of the phase space are given by eqs.(\ref{selec_limits}).
	    Also superimposed are the enclosing curves of {\em (a)}.
        } 
    \label{fig:selec_scatt}
\end{figure}

In actuality, such a  goal is much better
realized by examining the double 
differential cross section rather than making the individual 
cuts of eqs.(\ref{eta_e}--\ref{miss_mom}) any stronger. As the 
$W$-mediated diagrams have been suppressed by right-polarizing the 
electron beams, the bulk of the background owes its origin to 
the $e^- e^- Z$ final state with the $Z$ decaying invisibly. It is 
easy to see that for such a process, the energies of the two electrons 
satisfy the conditions~\cite{frank}
\be
\barr{rcl}
	E_1 + E_2  & \geq & \dis \frac{s - m_Z^2}{2 \sqrt{s}} 
		\\
	(2 E_1 - \sqrt{s}) (2 E_2 - \sqrt{s})  & \geq & m_Z^2 \ .
\earr
	\label{Z_bkgd_limits}
\ee
In Fig.~\ref{fig:selec_scatt}$a$, we present a scatter plot of 
the SM background for an accumulated luminosity of $50 \fb^{-1}$. 
Superimposed on it are the two curves of eq.(\ref{Z_bkgd_limits}). 
Eliminating the part of the phase space bounded by 
the two curves reduces the SM 
background from $ 19 \fb$ to approximately $1 fb$. The preponderance 
of points {\em just below} the straight line can be attributed to 
the contributions from a slightly off-shell $Z$.  A very large fraction 
of these could be eliminated by modifying the straight line 
curve by replacing $m_Z$ with, say, $m_Z - 2 \Gamma_Z$. Points 
well outside this region, on the other hand, 
owe their origin to the $W$-mediated diagrams and would disappear 
in the limit of fully right-polarized electron beams. 

In contrast to the above, the electrons in the signal events are 
distributed evenly  within the square region defined by 
\be 
\barr{rcl}
\dis	
  E_{\rm min} & \leq & E (e_i)  \leq  E_{\rm max} \ ,
	\\[2ex]
E_{\rm min, \ max} & = & \dis
\frac{\sqrt{s} }{4} \: 
	\left( 1 - \frac{m^2_{\widetilde G} }{m^2_{\widetilde e}} \right)
			\;   ( 1 \mp \beta ) \ ,
    \\[2ex]
\beta & = & \dis \left(1 - \frac{4 m_{\tilde e}^2}{s} \right)^{1/2} \ .
\earr
	\label{selec_limits}
\ee
This is illustrated in Fig.~\ref{fig:selec_scatt}$b$. While a 
significant fraction of the signal cross section could be lost 
on imposing the cuts corresponding to eq.(\ref{Z_bkgd_limits}), the 
signal to noise ratio shows an enormous improvement. 

\section{Neutralino as the NLSP}
	\label{sec:neut_NLSP}
As we have already pointed out, for $N_m = 1$, the lightest neutralino 
is always the NLSP. Even for $N_m > 1$, this may continue to be the case 
especially if the higgsino mass parameter $\mu$ is small. However, 
in all such cases the gaugino mass parameter $M_2$ is larger than the 
mass of the $\rslep$. Consequently, the selectron decays mainly into 
an electron and the lightest neutralino with the latter cascading 
into a photon and the gravitino. The final state thus comprises of a pair 
each of electrons and photons and missing energy due to the gravitinos. 
The energy and angular distributions would obviously depend on the 
masses of the selectron and the lightest neutralino. We will again 
concentrate on two representative points (see Table.~\ref{table:neut_cs})
in our analysis of the signal and comparison with the background.
\begin{table}[hb]
\begin{center}
\begin{tabular}{|r|r|r|r|r|r|r|r|r|}
\cline{2-9}
\multicolumn{1}{c|}{}
& \mult{\bf $M$} & \mult{\bf $\Lambda $} & 
	\mult{\bf $\mu$} & \mult{\bf $\tb$} &
   \mult{\bf $m_{\tilde e_{R}}$} & \mult{\bf $m_{\chi^0_1}$} 
	& \mult{{\bf $\sigma$} (fb)} & \mult{{\bf $\sigma $} (fb)}\\ 
\multicolumn{1}{c|}{}
& \multtev   &   \multtev   &  \multgev     
	&       & \multgev   &   \multgev
	& \mult{$\sqrt{s}=0.5$~TeV} & \mult{$\sqrt{s}=1$~TeV}\\
\hline
\pinegbf{(C)}& \pinegbf{140} & \pinegbf{70.0} & \pinegbf{--450} & 
	\pinegbf{3} & \pinegbf{126.6} & \pinegbf{102.7} & \pinegbf{1444} & 
		\pinegbf{277.1} \\
\hline
& 130  & 65.0 & 480 & 5 & 119.3 & 91.35 & 1372 & 267.3\\
\hline
& 150  & 75.0  & $-300$ & 6 & 135.9 & 108.1 & 1428 & 291.3 \\
\hline
& 160 & 80.0 & 400 & 6 & 144.2 & 112.1 & 1463 & 308.9 \\
\hline
& 200 & 100.0 & $-200$ & 8 & 178.0 & 135.4 & 1378 & 327.1 \\
\hline
& 240 & 80.0 & $-200$ &  9 &  145.4 & 107.6 & 1408 & 299.3 \\
\hline
& 230 & 115.0 & 250 & 9 & 203.7  & 153.5 & 1203 & 331.4 \\
\hline
& 275 & 137.5 & 350 & 10 & 242.4 & 193.2 & 571 & 340.0 \\
\hline
\mahogbf{(D)} & \mahogbf{500} & \mahogbf{125.0} & 
	\mahogbf{450} & \mahogbf{3} & \mahogbf{223.1} & \mahogbf{169.9} & 
	\mahogbf{1016} & \mahogbf{343.5} \\
\hline
\end{tabular}
\end{center}
\caption {Signal ($ e^- e^- + 2\gamma +  \mpT $ ) cross-section
          for some representative values of GMSB input parameters. The number 
	  of messenger generation $ N_m = 1$.
	  The cuts of eqs.(\protect\ref{eta_e}--\protect\ref{miss_mom}) and 
	  eq.(\protect\ref{photon_cuts}) have been imposed. 
	 }
	\label{table:neut_cs}
\end{table}


An exact calculation of the (6-body) SM background is an onerous task. 
It can be easily seen though that the bulk of the background arises from 
the two resonant processes: 
	\begin{enumerate}
	\item[$(i)$] $\emem \ra \emem \gamma \gamma Z \ra 
			  \emem \gamma \gamma \nu_i \bar \nu_i$ 
		\quad {\rm and}
	\item[$(ii)$] $\emem \ra e^- \bar \nu_e \gamma \gamma W^- \ra 
			  \emem \gamma \gamma \nu_e \bar \nu_e$ \ .
	\end{enumerate}
Once soft and collinear singularities have been removed by an 
appropriate set of cuts, one expects these cross sections to be 
smaller than those in the previous section by a factor ${\cal O}(\alpha^2)$. 
Thus the kinematical cuts required over and above 
those of eqs.(\ref{eta_e}--\ref{miss_mom}) are dictated not by 
the need to minimize background, but by detector acceptances.
To be specific, we demand that
\be
\barr{rclcrcl}
	|\eta_\gamma | & < & 3  & \quad & p_T(\gamma) & > & 10 \gev 
		\\
	\D R_{\gamma \gamma} & > & 0.2  & \quad 
		& \D R_{e \gamma} & > & 0.2  \ .
\earr
	\label{photon_cuts}
\ee
With these set of cuts, the surviving cross section, 
for $\sqrt{s} = 500 \gev$, from process $(i)$ 
above is approximately $0.004 \fb$ while that from the second one 
is approximately $0.001 \fb$. Thus, for all practical purposes, we 
have a background free situation. The surviving size of the signal 
is quite similar to that in the previous section as the cuts of 
eq.(\ref{photon_cuts}) do not take away much of the signal.

\begin{figure}[ht]
\centerline{
\epsfxsize=7cm\epsfysize=6cm
                     \epsfbox{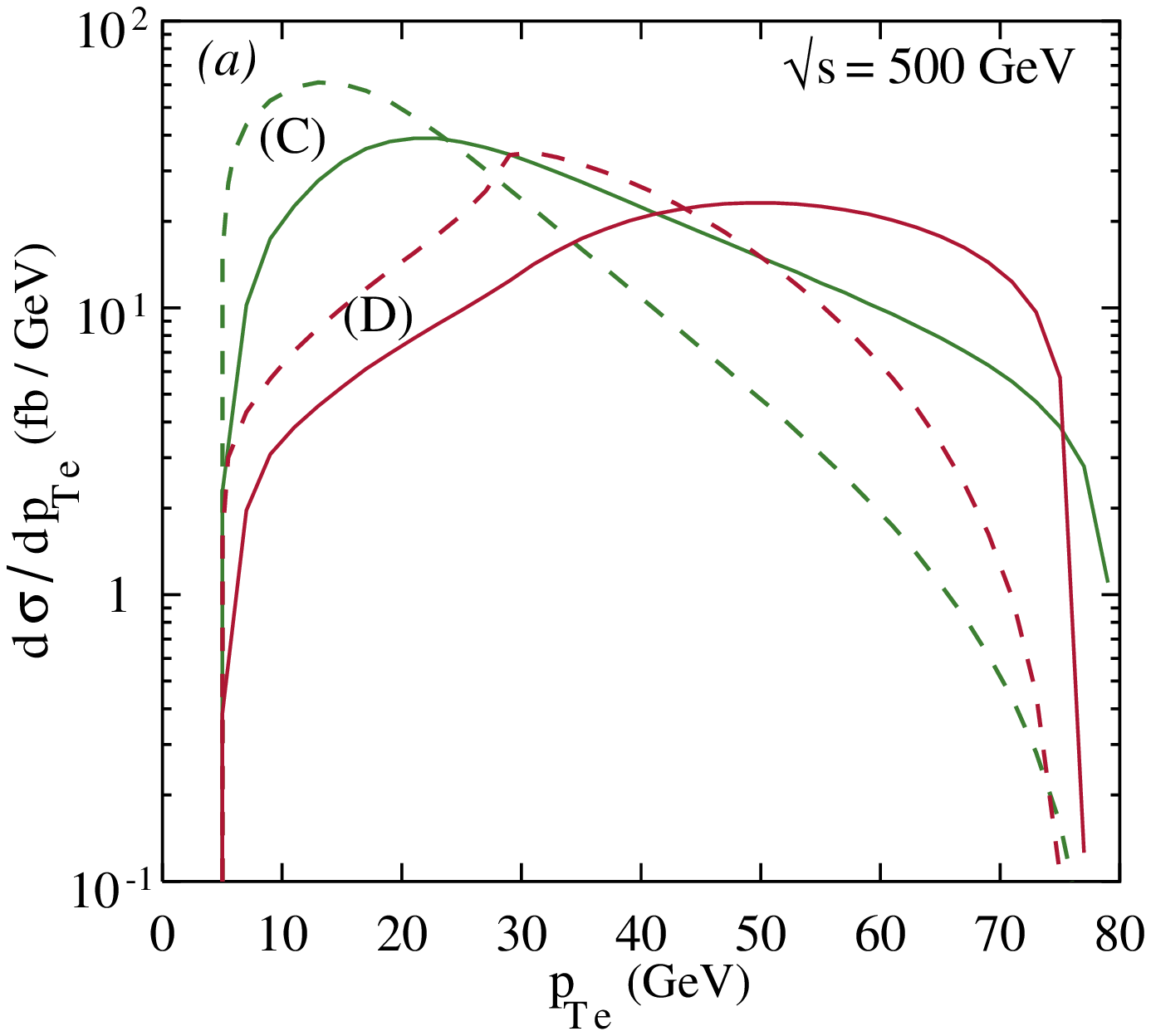}
\epsfxsize=7cm\epsfysize=6cm
                     \epsfbox{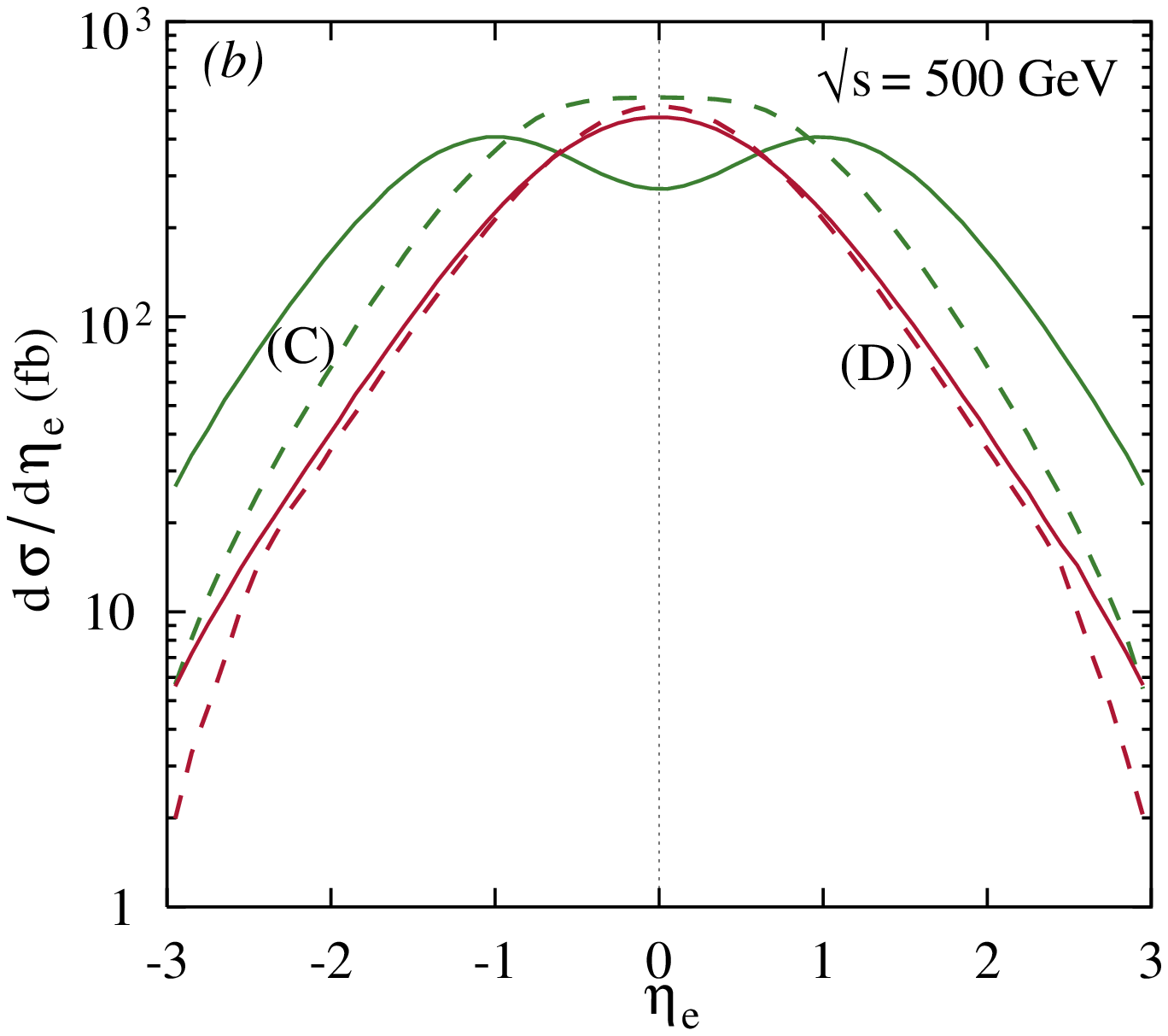}
}
\centerline{
\epsfxsize=7cm\epsfysize=6cm
                     \epsfbox{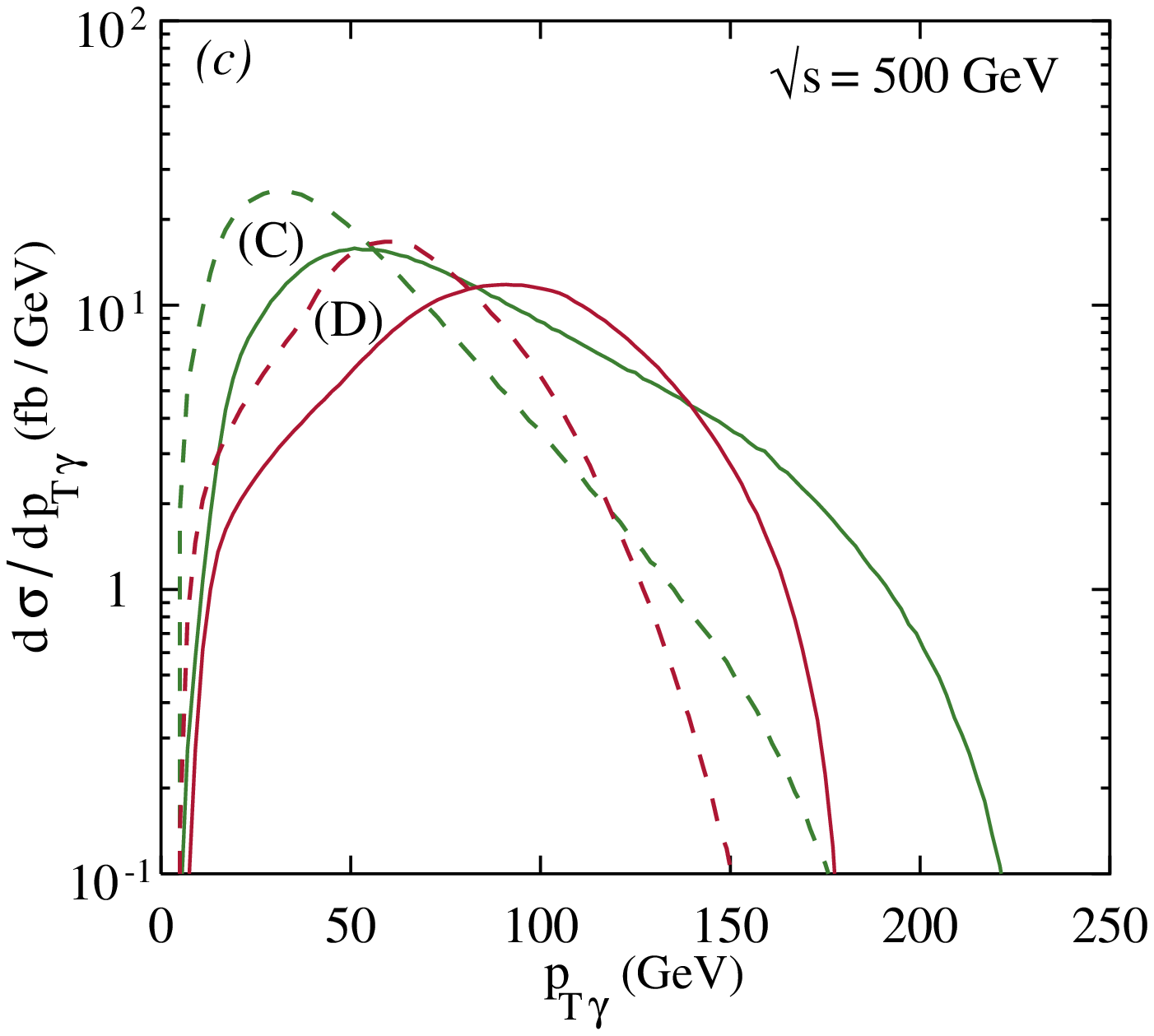}
\epsfxsize=7cm\epsfysize=6cm
                     \epsfbox{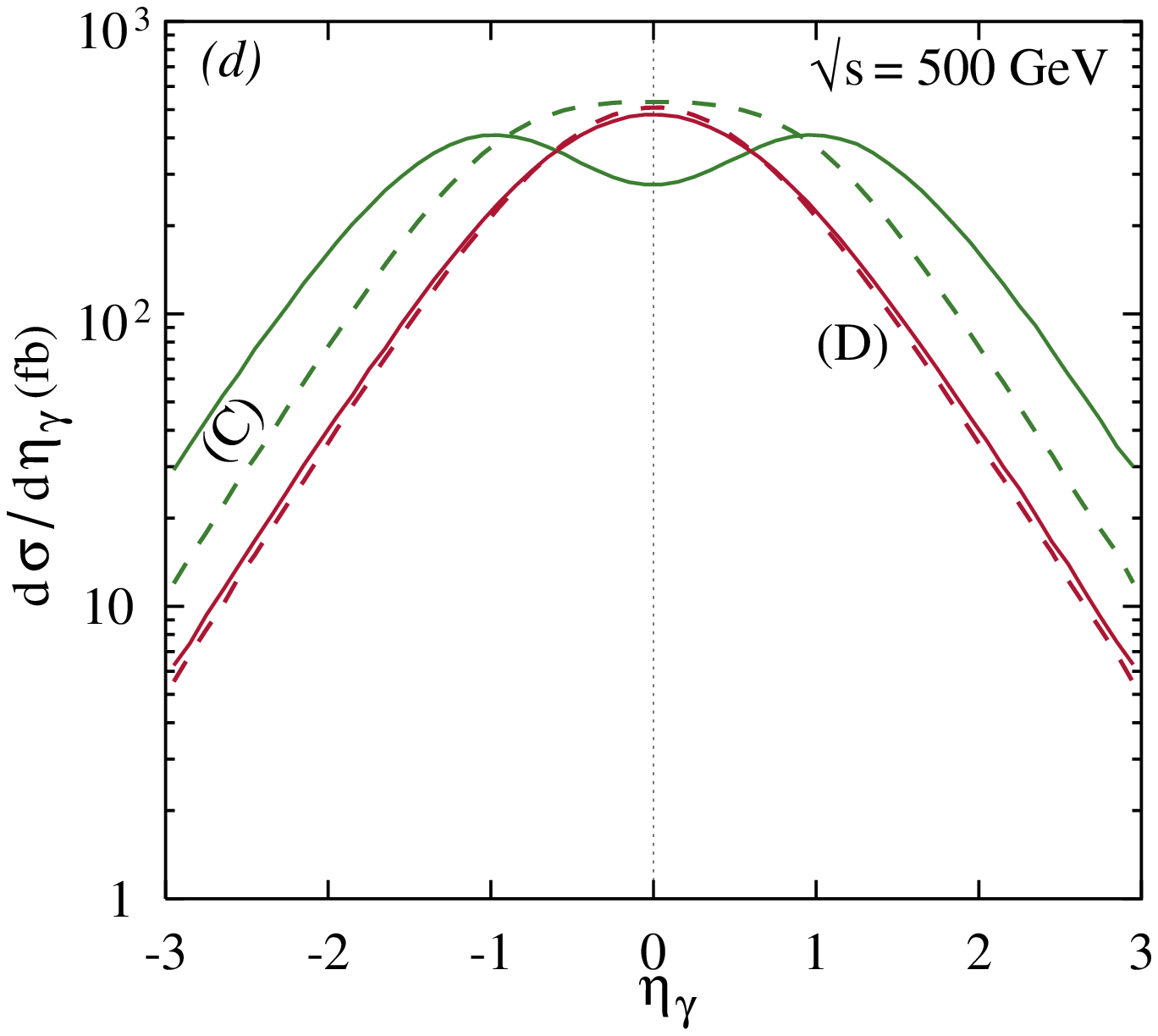}
}
\caption{\em Signal distributions for the neutralino NLSP case. The two 
	     sets correspond to parameters marked in 
     Table.~\protect\ref{table:neut_cs}. 
	     The SM background is too small to appear in the graphs
	     as the cuts of 
	     eqs.(\protect\ref{eta_e}--\protect\ref{miss_mom}) and 
	     eq.(\protect\ref{photon_cuts}) have been imposed.
	     Graph {\em (a)} pertains to
	     electron transverse momenta with the solid curves 
	     corresponding to the harder electron and the dashed to the 
	     softer one. Similarly, graph {\em (b)} pertains to the 
              electron rapidity distribution. {\em (c)} and {\em (d)} 
	     respectively give the  photon's transverse momenta and 
            rapidity distributions.
	}
    \label{fig:neut}
\end{figure}

In Fig.~\ref{fig:neut}, we present the signal event distributions 
for the two particular parameter choices indicated in 
Table.~\ref{table:neut_cs}. 
Comparing Figs.~\ref{fig:neut}$a$ and \ref{fig:selec}$a$, one is struck
by the similarity between the curves for parameter sets ($A$) and 
($C$) on the one hand and those for ($B$) and ($D$) on the other. 
This can be understood by realizing that the shape of electron transverse 
momenta distribution is determined by the masses of the selectron and the 
particles it is decaying into. Since cases ($A$) and ($C$) correspond 
to very similar $m_{\tilde e}$, it is only natural that the $p_T$ spectrum 
would look similar. 
The larger mass of the neutralino (as 
compared to that for the gravitino) is reflected in smaller value of 
the maximal $p_T$ allowed to the electrons. Analogous statements 
apply to points ($B$) and ($D$) as well. 

We turn now to the photon spectra. As the neutralinos 
are produced in (isotropic) scalar decays and as they themselves decay into 
a photon and gravitino, there are no nontrivial angular correlations. 
Consequently, the spectrum is determined by kinematics alone. This being 
identical to that for squark decay into massless particles, the decay 
distributions (Fig.~\ref{fig:neut}$c,d$) are very similar to those for 
the other set. A similar story obtains for the missing transverse momentum 
as well.

\section{Distinguishing from non-GMSB models}
	\label{sec:identify} 
In the last two sections we have seen that, for $\L \lsim 200 \tev$, 
the signal from a GMSB model stands well and truly above the SM background.
This is particularly true for the neutralino NLSP case where one 
expects less than one event from SM processes. This brings us to the 
more important question, namely how to recognize if supersymmetry 
breaking is driven by gauge mediation. It goes without
saying, though, that short of determining the 
entire spectrum, one can only draw (strong) inferences and this is what 
we shall aim to do in this section. 

\begin{figure}[ht]
\centerline{
\epsfxsize=9cm\epsfysize=8cm
                     \epsfbox{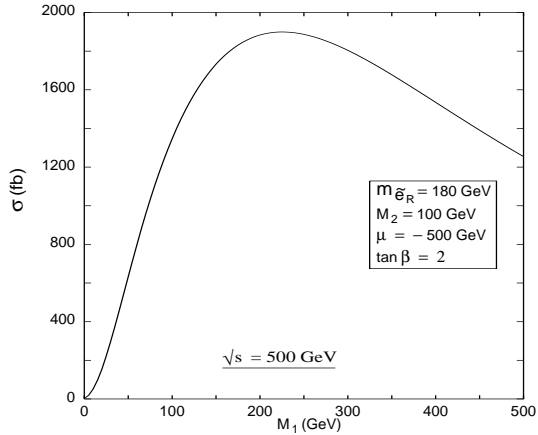}
	   }
\caption{\em The variation of $\rslep$ pair production 
	     cross section with the gaugino mass parameter $M_1$.
	     No assumption regarding SUSY breaking has been made.
	     The dependence on $M_2$, $\mu$ and $\tan \beta$ 
	     is negligible.
        } 
       \label{fig:light_bino}
\end{figure}

Considering the selectron NLSP case first, it is clear that 
eq.(\ref{selec_limits}) can be used to determine the masses 
of both the selectron and the supersymmetric particle $X$ (gravitino
or neutralino) that it is decaying into. Of course, measurement of the 
edge of phasespace is always beset with inaccuracies. However, at this 
stage we do not need to know the mass of $X$ very accurately. In fact,
as long as the experimentally deduced value 
$m_X \lsim 30 \gev$ or so, the rest of argument follows. LEP data 
already tells us that such a light neutralino can only be the bino 
(primarily)\footnote{Unless there exist neutralinos, and hence gauge 
	symmetries, going beyond the MSSM. We do not consider 
	such exotic models}. 
Now, the $\tilde e_R$ does not couple with the $\tilde W_3$ and its 
coupling with the higgsinos is suppressed by the electron mass. Consequently,
for a $\tilde e_R$ of fixed mass,
the production cross section (eq.\ref{prod_cs}) is determined essentially
by the bino mass $M_1$. Working in the limit $M_2, \mu \gg M_1$, 
the chirality structure of the amplitude 
ensures that, for small values of $M_1^2 / s$, the cross section 
grows as this ratio (see Fig.~\ref{fig:light_bino}).  
For large values of the ratio, though, the cross 
section would fall off. Realistic values for $M_2$ and $\mu$ would 
alter our simplistic arguments to a degree, but such effects are too 
small to be noticeable in the graph that we present. 

That we have produced a pair of $\tilde e_R$s and not $\tilde e_L$s 
we can deduce from the polarization of the initial state. Its mass,
as we have already seen, can be determined from the energy 
distribution, and if necessary, refined by a threshold scan. 
At this stage, Fig.~\ref{fig:light_bino} can be used to ``determine'' 
$M_1$ from the experimentally measured cross section and compare 
it with the direct, if inaccurate, measurement from the endpoint 
analysis. Clearly, the consistency between the two values 
would be much higher for the GMSB hypothsis than for the 
non-GMSB case. Thus, rate counting helps us to 
to distinguish between the light gravitino and light bino cases.

The neutralino NLSP case presents us with an additional complication. 
Presumably we could have had $\rslep \ra e^- + \chi_2^0$ in a
non-GMSB scenario followed by $\chi_2^0 \ra \chi_1^0+ \gamma$. We can 
again measure both $m_{\rslep}$ and $m_{\chi}$ by determining 
the phasespace boundaries of the electrons and employing a relation 
analogous to eq.(\ref{selec_limits}). What about the mass of the 
gravitino (neutralino) in the second stage of the decay? In fact, a
corresponding relation can be derived for the energy of the photons:
\be
\barr{rcl}
    E_\gamma^{\rm max, min} & = & \dis A \: (1 \pm \beta) \: (1 \pm \zeta)
	\\[2ex]
     A  & \equiv & \dis 
	\frac{ \sqrt{s} }{8}   
         \left( 1 + \frac{m_\chi^2}{m^2_{\se}}  \right)
         \: \left( 1 - \frac{m^2_{\grav}}{m_\chi^2} \right)
	\\[2ex]
      \zeta & \equiv & \dis \frac{  m_\chi^2 - m^2_{\grav} }
                   {  m_\chi^2 + m^2_{\grav} } \ ,
\earr
	\label{photon_limits}
\ee
with $\beta$ defined as in eq.(\ref{selec_limits}). 
\begin{figure}[ht]
\centerline{
\epsfxsize=7cm\epsfysize=6cm
                     \epsfbox{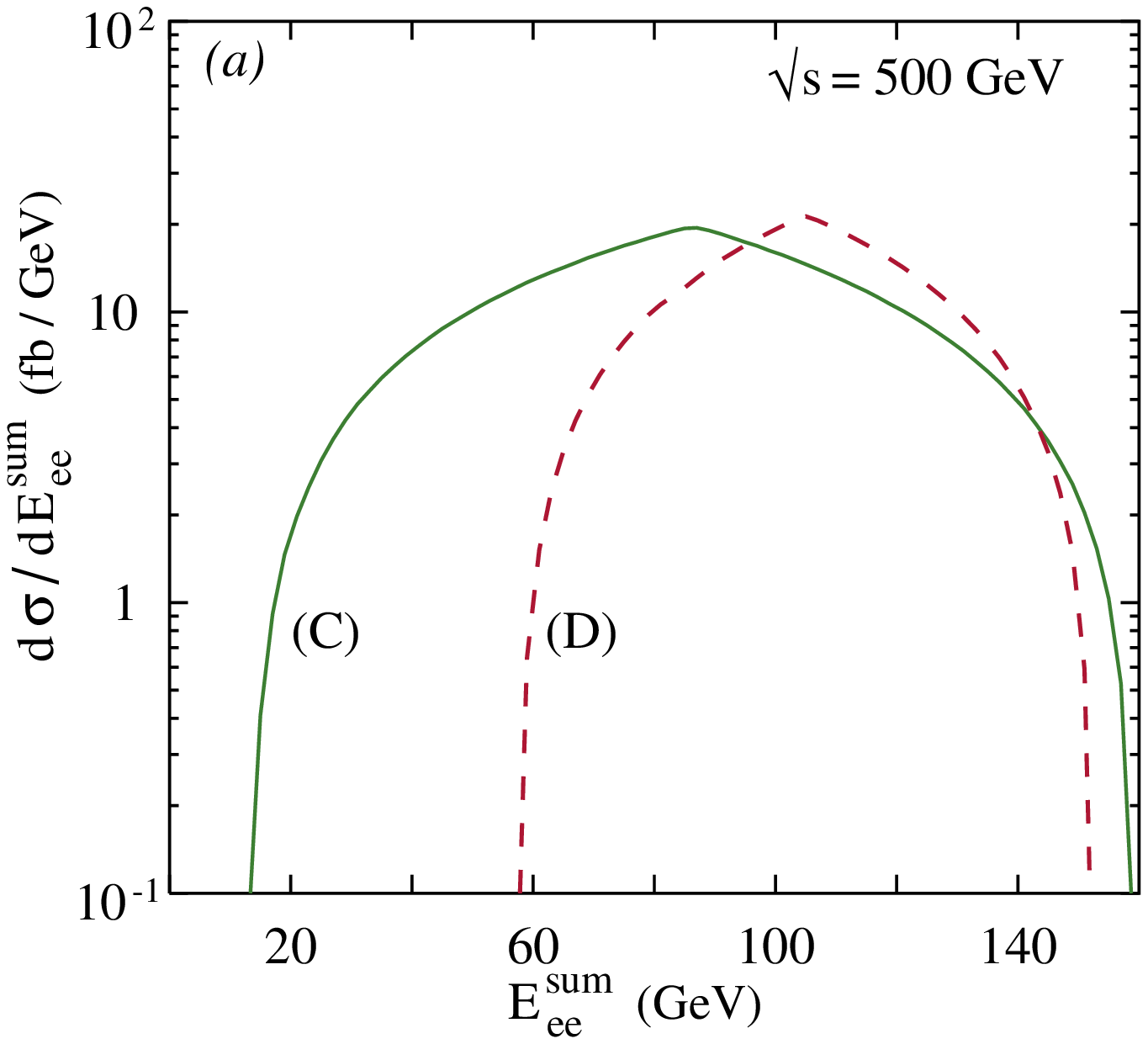}
\epsfxsize=7cm\epsfysize=6cm
                     \epsfbox{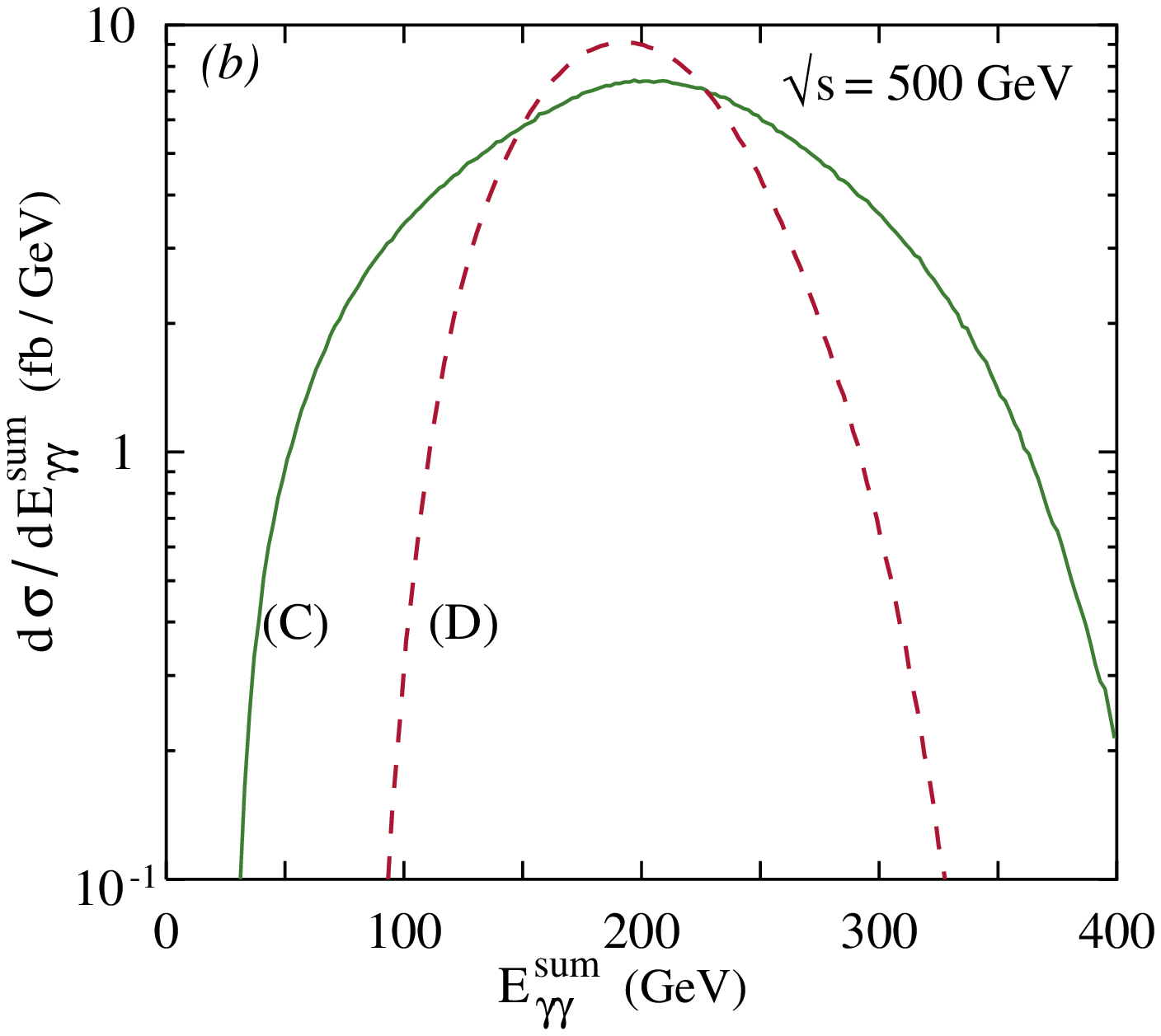}
}

\centerline{
\epsfxsize=7cm\epsfysize=6cm
                     \epsfbox{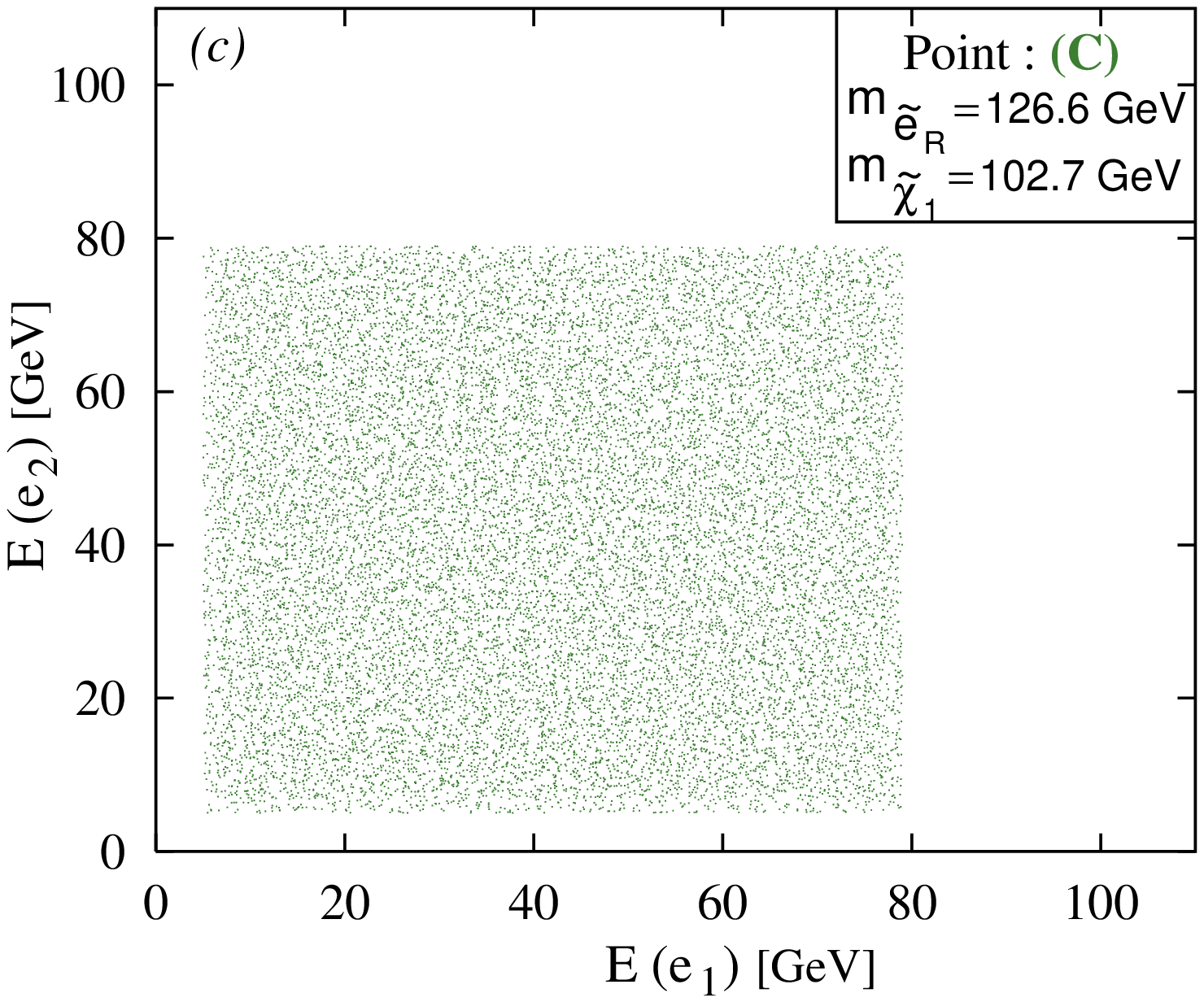}
\epsfxsize=7cm\epsfysize=6cm
                     \epsfbox{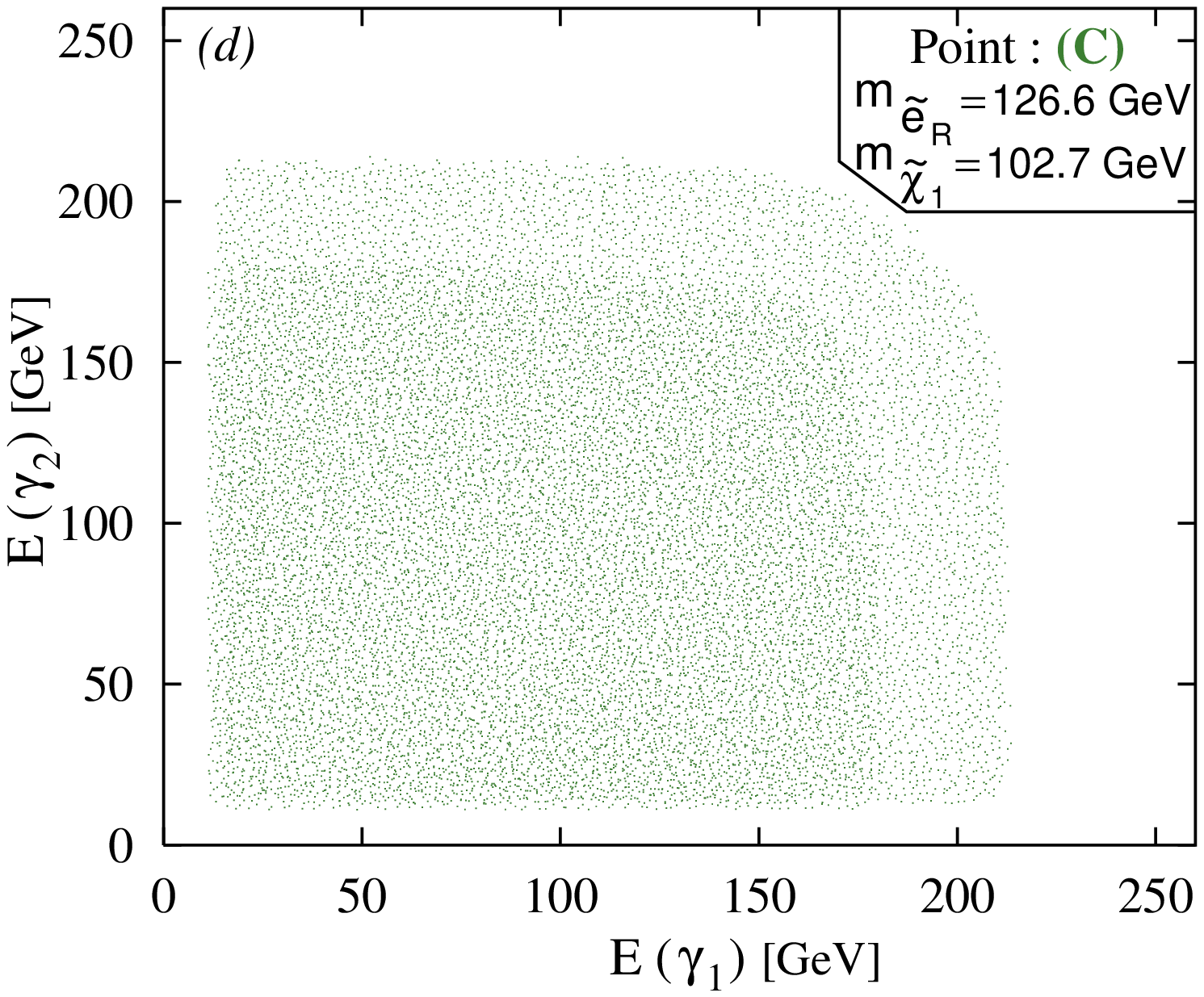}
}
\caption{\em {\em (a)} The distribution in the scalar sum of the electron 
	   energies for the signal events in the neutralino NLSP case.	   
	   The cuts of eqs.(\protect\ref{eta_e}--\protect\ref{miss_mom}) and 
	   eq.(\protect\ref{photon_cuts}) have been imposed. The legends 
	   refer to the parameter set marked in  
	   Table.~\protect\ref{table:neut_cs}. THE SM background
	   is too small to appear on the graph.
	   {\em (b)} Similar distribution but in the scalar sum of 
	   photon energies. 
	   {\em (c)} Scatter plot in the electron energies for an integrated 
	   luminosity of $50 \fb^{-1}$. Each point corresponds to 
	   {\bf four} events. 
	   {\em (d)} Similar scatter plot in the photon energies.
	}
    \label{fig:neut_scatt}
\end{figure}

In Figs.~\ref{fig:neut_scatt}($a,b$), we exhibit the distributions in the 
scalar sums of electron and photon energies respectively. The 
first, namely $E_{ee}^{\rm sum} \equiv E_{e_1} + E_{e_2}$, 
leads to a symmetric distribution as in the case of a selectron NLSP. 
On the other hand, the distribution in $E_{\gamma\gamma}^{\rm sum}$ shows a 
high energy tail. The tail is purely a kinematic feature and can 
be derived from a generalization of the Dalitz plot. A better understanding 
of the same can be obtained from the scatter plots of 
Figs.~\ref{fig:neut_scatt}($c,d$). 

As in the previous case, the endpoints of the electron spectrum
can be used to deduce both $m_{\tilde e_R}$ and $m_\chi$. 
Once $m_{\tilde e_R}$ is measured, eq.(\ref{photon_limits}) can 
be used to determine both $m_\chi$ and $m_{\grav}$. 
This, thus, also serves as a consistency check. At this stage we can again 
take recourse to Fig.~\ref{fig:light_bino} to argue that the existence 
of such a light bino (as in a non-GMSB model) 
would have implied a small cross section. Moreover, 
if such a bino were to exist, the selectron would have a substantial 
branching into it. Hence, nonobservance of an excess in the 
($\emem +$ missing energy) final state is yet another argument against 
a spectrum with a heavy gravitino but a light bino.
 
\section{Summary}
	\label{sec:concl}
The $e^-e^-$ option of the Next Generation Linear collider
can be a very effective tool in the search for physics beyond the SM.
In this paper, we have studied the feasibility of using such a machine 
to probe Gauge Mediated Supersymmetry Breaking. The process of choice is 
the pair-production of right-handed selectrons, not in the least because 
of their being significantly lighter than their left-handed 
counterparts. 

If the selectron be the NLSP, the signal comprises two electrons accompanied 
by a missing momentum. Right-polarizing (90\%) the electron beams helps 
eliminate the bulk of the SM background apart from increasing the signal 
strength as well. The already very good signal to noise ratio can be enhanced 
even further by imposing correlated cuts on the electron energies. 
The neutralino NLSP case, on the other hand results in a spectacular 
final state comprising a pair each of electrons and photons accompanied by
missing momentum. The SM background is virtually nonexistent. In either case,
the rates are high enough for the selectron to be detectible almost upto
the kinematic limit. 

The energy correlations (electrons for the selectron NLSP case and 
both electrons and photons in the neutralino NLSP case) are characteristic
and can be used to determine the masses of both the produced particle
and its decay products. Furthermore, such information gleaned from the
differential distributions, used in conjunction with rate counts, 
can be used to distingguish GMSB from alternate scenarios 
of supersymmetry breaking (including, but not limited to, 
the case of supergravity-inspired models without gaugino mass 
unification).

\vskip 25pt
\begin{center}
{\bf Acknowledgement }
\end{center}
DC acknowledges the Department of Science and Technology, India 
for the Swarnajayanti Fellowship grant. 
DKG acknowledges the hospitality of the Theory Divison, CERN, 
Geneva and the Laboratorie de Physique Particles (LAPP), Annecy, where a
part of this work was done.
\newpage

\newcommand{\ib}[3]   {{\em ibid.\/} {\bf #1} (#3) #2}		       %
\newcommand{\app}[3]  {{\em Acta Phys. Polon.	B\/}{\bf #1} (#3) #2}  %
\newcommand{\ajp}[3]  {{\em Am. J. Phys.\/} {\bf #1} (#3) #2}	       %
\newcommand{\ap}[3]   {{\em Ann. Phys.	(NY)\/}	{\bf #1} (#3)	#2}    %
\newcommand{\araa}[3] {{\em Annu. Rev. Astron. Astrophys.\/}	       %
          {\bf#1} (#3) #2}					       %
\newcommand{\apj}[3]  {{\em Astrophys. J.\/} {\bf #1} (#3) #2}         %
\newcommand{\apjs}[3] {{\em Astrophys. J. Suppl.\/}                    %
          {\bf	#1} (#3) #2}			                       %
\newcommand{\apjl}[3] {{\em Astrophys. J. Lett.\/} {\bf #1} (#3) #2}   %
\newcommand{\astropp}[3]{Astropart. Phys. {\bf #1} (#3) #2}	       %
\newcommand{\eur}[3]  {Eur. Phys. J. {\bf C#1} (#3) #2}                %
\newcommand{\iauc}[4] {{\em IAU Circular\/} #1                         %
       (\ifcase#2\or January \or February \or March  \or April \or May %
                 \or June    \or July     \or August \or September     %
                 \or October \or November \or December                 %
        \fi \ #3, #4)}					               %
\newcommand{\ijmp}[3] {Int. J. Mod. Phys. {\bf A#1} (#3) #2}           %
\newcommand{\jetp}[6] {{\em Zh. Eksp. Teor. Fiz.\/} {\bf #1} (#3) #2   %
     [English translation: {\it Sov. Phys.--JETP } {\bf #4} (#6) #5]}  %
\newcommand{\jetpl}[6]{{\em ZhETF Pis'ma\/} {\bf #1} (#3) #2           %
     [English translation: {\it JETP Lett.\/} {\bf #4} (#6) #5]}       %
\newcommand{\jhep}[3] {JHEP {\bf #1} (#3) #2}                          %
\newcommand{\mpla}[3] {Mod. Phys. Lett. {\bf A#1} (#3) #2}             %
\newcommand{\nat}[3]  {Nature (London) {\bf #1} (#3) #2}	       %
\newcommand{\nuovocim}[3]{Nuovo Cim. {\bf #1} (#3) #2}	               %
\newcommand{\np}[3]   {Nucl. Phys. {\bf B#1} (#3) #2}		       %
\newcommand{\npbps}[3]{Nucl. Phys. B (Proc. Suppl.)                    %
           {\bf #1} (#3) #2}	                                       %
\newcommand{\philt}[3] {Phil. Trans. Roy. Soc. London A {\bf #1} #2    %
	(#3)}							       %
\newcommand{\prev}[3] {Phys. Rev. {\bf #1} (#3) #2}	       	       %
\newcommand{\prd}[3]  {Phys. Rev. {\bf D#1} (#3) #2}		       %
\newcommand{\prl}[3]  {Phys. Rev. Lett. {\bf #1} (#3) #2}	       %
\newcommand{\plb}[3]  {{Phys. Lett.} {\bf B#1} (#3) #2}		       %
\newcommand{\prep}[3] {Phys. Rep. {\bf #1} (#3) #2}		       %
\newcommand{\ptp}[3]  {{\em Prog. Theoret. Phys. (Kyoto)\/}            %
          {\bf #1} (#3) #2}					       %
\newcommand{\rpp}[3]  {Rep. Prog. Phys. {\bf #1} (#3) #2}              %
\newcommand{\rmp}[3]  {Rev. Mod. Phys. {\bf #1} (#3) #2}               %
\newcommand{\sci}[3]  {Science {\bf #1} (#3) #2}		       %
\newcommand{\zp}[3]   {Z.~Phys. C{\bf#1} (#3) #2}		       %
\newcommand{\uspekhi}[6]{{\em Usp. Fiz. Nauk.\/} {\bf #1} (#3) #2      %
     [English translation: {\it Sov. Phys. Usp.\/} {\bf #4} (#6) #5]}  %
\newcommand{\yadfiz}[4]{Yad. Fiz. {\bf #1} (#3) #2 [English	       %
	transl.: Sov. J. Nucl.	Phys. {\bf #1} #3 (#4)]}	       %
\newcommand{\hepph}[1] {(electronic archive:	hep--ph/#1)}	       %
\newcommand{\hepex}[1] {(electronic archive:	hep--ex/#1)}	       %
\newcommand{\astro}[1] {(electronic archive:	astro--ph/#1)}	       %

\end{document}